\documentclass[pre, twocolumn, preprintnumbers, amsmath, amssymb, nofootinbib, floatfix, superscriptaddress]{revtex4}

\usepackage{graphicx, bm, xcolor, bbold}
\usepackage{enumerate}
\usepackage[normalem]{ulem} 
\usepackage[breaklinks]{hyperref}

\makeatletter
\def\graphicscale{\twocolumn@sw{0.3}{0.4}}
\def\graphicthreescale{\twocolumn@sw{0.3}{0.4}}

\begin{document}

\title{Dynamic scaling behaviors induced by periodic drivings \\
  at continuous Ising quantum transitions}

\author{Andrea Pelissetto}
\altaffiliation{Authors are listed in alphabetic order.}
\affiliation{Dipartimento di Fisica dell'Universit\`a di Roma
  La Sapienza and INFN, Sezione di Roma I, I-00185 Roma, Italy}

\author{Davide Rossini}
\altaffiliation{Authors are listed in alphabetic order.}
\affiliation{Dipartimento di Fisica dell'Universit\`a di Pisa
  and INFN, Largo Pontecorvo 3, I-56127 Pisa, Italy}

\author{Ettore Vicari}
\altaffiliation{Authors are listed in alphabetic order.}
\affiliation{Dipartimento di Fisica dell'Universit\`a di Pisa,
  Largo Pontecorvo 3, I-56127 Pisa, Italy}

\date{\today}

\begin{abstract}
  We investigate the out-of-equilibrium scaling behaviors induced by
  periodic driving protocols in quantum many-body systems near
  continuous quantum phase transitions at zero temperature. We
  consider two paradigmatic one-dimensional systems: the quantum Ising
  chain subject to periodic longitudinal fields and the Kitaev
  fermionic wire with periodically driven chemical potentials,
  equivalent to a quantum Ising chain with a periodic transverse
  field. Within a renormalization-group (RG) framework, we develop a
  dynamic scaling theory for both finite-size systems and the
  thermodynamic limit, identifying scaling variables associated with
  the driving amplitude and period.  We distinguish between protocols
  that break the global ${\mathbb Z}_2$ symmetry by coupling to the
  order parameter and those that preserve it.  These two classes of
  driving exhibit distinct scaling behaviors, with power-law exponents
  determined by the RG dimensions of the corresponding
  perturbations. We verify the proposed dynamic scaling theory through
  numerical simulations of both smooth single-frequency and
  discontinuous square-wave periodic drivings.
\end{abstract}

\maketitle

\section{Introduction}
\label{intro}

Critical phenomena at classical and quantum phase transitions exhibit
a variety of equilibrium and nonequilibrium scaling behaviors (see,
e.g.,
Refs.~\cite{Wilson-83,WK-74,Fisher-74,Wegner-76,Aharony-76,Ma-book,
  Binder-76,HH-77,Kibble-80,Zurek-96,Cardy-book,SGCS-97,PV-02,CG-05,FM-06,
  Sachdev-book,PSSV-11,Biroli-16,RV-21,BPV-25} for a nonexhaustive
list of books and reviews).  Non-equilibrium dynamics driven by
time-dependent system parameters can also give rise to interesting
scaling behaviors.  A notable example is the nonequilibrium
Kibble–Zurek dynamics~\cite{Kibble-80,Zurek-96}, in which a relevant
parameter is slowly driven across a phase transition
~\cite{CEGS-12,RV-21}.  Other interesting dynamical protocols involve
periodic driving, in which one or more system parameters vary
cyclically in time. Such protocols may give rise to distinctive
nonequilibrium scaling behaviors near continuous phase transitions.
They have already been studied in the classical Ising model, under
periodic driving of the temperature and of the external magnetic field
across the critical point~\cite{PV-26-per}.  They show power-law
scaling behavior depending on the critical exponents of the Ising and
relaxational dynamics universality classes, and their nonequilibrium
scaling functions appear to be characterized by long-time stationary
behavior and synchronization. Other studies have also investigated the
effects of periodic driving in ferromagnetic systems, focusing on
nonequilibrium dynamical transitions that are distinct from the
underlying equilibrium transition (see, e.g.,
Refs.~\cite{LP-90,SRN-98,CA-99,PP-13,MBR-24,BMR-25,XKM-26}).

Periodic driving has been extensively studied in quantum many-body
systems evolving under unitary dynamics (see, e.g.,
Refs.~\cite{KIMG-06,GD-14,RRDOR-23,CUBZL-23,SG-26,YTZL-26,MSB-26,
  BDD-12,RSS-12,LMPPA-17,CZQXX-25,GPJ-26}), often within the Floquet
framework for time-periodic quantum Hamiltonians (see, e.g., the
review articles~\cite{GH-98,SAN-10,WBDPVK-17,SSS-21} and references
therein).  Some of these studies focus on the paradigmatic quantum
Ising chain subject to a periodic time-dependent transverse
field~\cite{BDD-12,RSS-12,LMPPA-17,CZQXX-25,GPJ-26}.

In this paper, we investigate the finite-size and thermodynamic-limit
scaling behaviors induced by periodic driving in the critical regime
of continuous quantum phase transitions. Unlike classical systems
governed by a dissipative relaxational dynamics~\cite{Ma-book}, closed
quantum systems evolve unitarily and are therefore
nondissipative. This fundamental difference is expected to give rise
to dynamical behaviors that differ substantially from those observed
in dissipative classical systems ~\cite{PV-26-per}.

We address these issues by studying two paradigmatic one-dimensional
quantum many-body systems: the quantum Ising chain and the Kitaev
fermionic wire, which are related by the nonlocal Jordan–Wigner
transformation~\cite{Sachdev-book}. We investigate the scaling
behavior induced by periodically driving either the longitudinal or
transverse homogeneous magnetic field across the quantum Ising-chain
critical point, as well as the chemical potential across the critical
transition in the Kitaev fermionic wire. We analyze these behaviors
within the renormalization-group (RG) framework, considering both the
finite-size scaling (FSS) and the scaling in the thermodynamic
limit. Our scaling arguments are quite general and apply to continuous
quantum phase transitions beyond the specific models considered
here. We then test these predictions numerically on one-dimensional
Ising chains and Kitaev fermionic wires.

The paper is organized as follows. In Sec.~\ref{quischprot}, we
introduce the quantum Ising model in the presence of external
transverse and longitudinal fields, and outline the dynamic protocols
realizing the periodic driving of the fields.
In Sec.~\ref{scatheo}, we develop a dynamic scaling theory
for the scaling phenomena induced by periodic driving protocols coupled to the
order-parameter operator associated with quantum Ising transitions,
and verify the dynamic FSS predictions through numerical simulations
of quantum Ising chains with periodic longitudinal field, focusing on
square-wave driving protocols.  In Sec.~\ref{traperkit}, we
investigate the critical effects of periodic variations of the
chemical potential in Kitaev fermionic wires, which are equivalent to
quantum Ising chains with periodically driven transverse fields. We
describe the corresponding dynamic scaling theory and numerically
verify the scaling predictions in both the FSS and thermodynamic
limits, considering smooth single-frequency and discontinuous
square-wave driving protocols.  Finally, in Sec.~\ref{conclu}, we summarize 
the results, draw our conclusions, and discuss possible future
developments. In App.~\ref{endensque}, we present analytical
calculations showing how logarithmic contributions invalidate
the proposed pure power-law scaling of the excess energy
under square-wave driving.

\section{Quantum Ising chains with periodic driving}
\label{quischprot}

\subsection{The model}
\label{model}

We consider the nearest-neighbor one-dimensional quantum Ising model
with Hamiltonian
\begin{equation}
  \hat H(g,h) = - \sum_{x=1}^L \left[ J\,\hat\sigma^{(1)}_x
    \hat\sigma^{(1)}_{x+1} + g \, \hat\sigma^{(3)}_x + h
    \,\hat\sigma^{(1)}_x\right] ,
  \label{hedef}
\end{equation}
where $\hat\sigma_x^{(k)}$ are Pauli matrices ($k=1,2,3$) defined on
the $x$th lattice site. The Hamiltonian parameters $g$ and $h$ play
the role of transverse and longitudinal homogeneous magnetic fields. We
consider chains with $L$ sites and periodic boundary conditions
($\hat\sigma^{(k)}_x=\hat\sigma^{(k)}_{L+x}$), preserving
translational invariance (no boundaries are present) and the ${\mathbb
  Z}_2$ symmetry when $h=0$.  Without loss of generality, we assume $J
= 1$ and $g>0$. We also set $\hbar = 1$.

The quantum Ising chain (\ref{hedef}) is a paradigmatic
system~\cite{Sachdev-book,RV-21} undergoing a continuous quantum
transition at $g=g_c=1$ and $h=0$, belonging to the 
two-dimensional Ising universality class. It separates a
disordered ($g>g_c$) from an ordered phase ($g<g_c$). 
For $g<g_c$ the model exhibits a magnetic first-order
quantum transition line, driven by the longitudinal field
and characterized by the discontinuity of the longitudinal
magnetization.  Approaching the continuous transition, the
system develops long-distance correlations, with length scale $\xi$
diverging as
\begin{equation}
  \xi \sim |r|^{-\nu},\quad  r=g-g_c,\quad \nu=y_r^{-1}=1,
  \label{xirdef}
  \end{equation}
where $y_r$ is the RG dimension of the deviation $r$ from the critical
point $g_c=1$. In the same regime
the energy gap between the lowest-energy states behaves 
as $\Delta \sim \xi^{-z}$, where $z$ is the dynamic critical
exponent; for Ising systems we have $z=1$. 
Another independent critical exponent is related 
to the RG dimension $y_h$ of the symmetry-breaking longitudinal
field $h$, 
\begin{equation}
  y_h={2+d+z-\eta\over 2} = 2 - \frac{\eta}{2} = \frac{15}{8},\qquad
  \eta=\frac{1}{4},
  \label{yhdef}
\end{equation}  
where $d$ stands for the system dimension (here $d=1$).
The RG dimension $y_\phi$ of the 
spin operator $\hat \sigma^{(1)}_x$ is
\begin{equation}
  y_\phi = d + z - y_h = \frac{\eta}{2} = \frac{1}{8}.
  \label{yphival}
\end{equation}
Finally, the RG dimension $y_n$ of the transverse spin operator
$\hat\sigma^{(3)}_x$ is given by
\begin{equation}
  y_n = d + z - y_r = 1.
\label{ynval}
\end{equation}  
We investigate the out-of-equilibrium behavior induced by periodic
time-dependent driving across the continuous quantum phase transition
at $g=g_c$ and $h=0$. 
Periodic driving protocols fall into two main classes 
depending on whether they preserve or break the global $\mathbb{Z}_2$ symmetry.
Periodic variations $h(t)$ of the longitudinal field around $h=0$ break the 
symmetry, whereas
periodic variations $g(t)$ of the transverse field preserve it. In the
following, we discuss the two cases separately.

\subsection{Periodic longitudinal driving}
\label{dynscaloper}

We now outline a dynamic protocol to study the quantum unitary
dynamics driven by a periodic longitudinal field $h(t)$ around $h=0$,
keeping the transverse field $g$ fixed within the critical regime
(i.e., $g\approx g_c=1$).  The periodic dynamic protocol starts
at $t=0$ from the quantum many-body state
$|\Psi(t=0)\rangle=|\Phi_0(g,h_i)\rangle$, where $|\Phi_0(g,h)\rangle$
is the ground state of the Hamiltonian~(\ref{hedef}).
The evolution of the system is described by the density matrix
\begin{equation}
  \hat\rho(t) = |\Psi(t) \rangle \langle\Psi(t)|,
  \label{rhot}
\end{equation}
where $|\Psi(t) \rangle$ is the solution of the Schr\"odinger equation
\begin{subequations}
  \begin{eqnarray}
    &&i \partial_t |\Psi(t) \rangle = \hat H[g,h(t)] |\Psi(t) \rangle, 
    \label{scheq}\\
    &&|\Psi(t=0) \rangle = |\Phi_0(g,h_i) \rangle,
  \end{eqnarray}
\end{subequations}
with a periodic longitudinal field $h(t)$.

We consider the single-frequency protocol with 
\begin{equation}
  h(t) = A \cos(\omega t), \quad \omega = \frac{2\pi}{P},\qquad h_i=A,
  \label{hstcos}
\end{equation}
where $A$ and $P$ are the amplitude and the period of the modulation,
respectively, and the computationally more convenient
square-wave protocol with 
\begin{equation}
  h(t) = (-1)^{n(t)} A, \quad n(t) = \lceil 2\,t/P \rceil \in {\mathbb N}, \quad
  h_i=A,
\label{hst}
\end{equation}
where $A>0$ and $\lceil x\rceil = {\rm min}\{ n\in {\mathbb Z} | n\ge
x\}$ is the ceiling function. Again, $A$ and $P$ are the amplitude and
period of the driving longitudinal field $h(t)$, respectively.  Note
that for $0 < t < P/2$ this protocol is equivalent to a standard quench,
in which the longitudinal field $h$ is instantaneously changed to
$h=-A$ at $t=0$, starting from the ground state for $h_i=A$.
The initial state is prepared before the sudden change of
the field at $t = 0$. At each switching time, the state remains
continuous, while the Hamiltonian changes discontinuously.
For observables depending explicitly on the instantaneous field,
the values reported at switching times refer to the right-hand limit,
immediately after the field change. The initial ground-state values
at $t = 0$ are distinguished from the post-quench values at $t = 0^+$.

The quantum evolution for $t>0$ can be monitored by evaluating the
instantaneous expectation value of the longitudinal magnetization
operator,
\begin{equation}
  M(t) = {\rm Tr} \big[ \hat\rho(t) \, \hat M \big],\qquad
\hat M \equiv \frac{1}{L} \sum_x \hat\sigma_x^{(1)},
  \label{mxm}
\end{equation}
and that of its square operator,
\begin{equation}
  M_2(t) = {\rm Tr} \big[ \hat\rho(t) \, \hat M^2 \big].
  \label{mxm2}
\end{equation}
We also consider the instantaneous expectation value of the transverse
magnetization operator,
  \begin{equation}
    U(t) = {\rm Tr} \big[ \hat\rho(t) \, \hat U \big],
    \qquad \hat U
  \equiv \frac{1}{L} \sum_x \hat\sigma_x^{(3)},
  \label{mxmz}
\end{equation}
and the related subtracted quantity
\begin{eqnarray}
  U_s(t) &=& U(t) -  U_0(g,L), \label{submmxmz}\\
  U_0(g,L) &=&
  \langle\Phi_0(g,h=0)| \hat U | \Phi_0(g,h=0) \rangle.\label{neqdef}
\end{eqnarray}
Here $U_0$ is the ground-state transverse magnetization
at zero longitudinal field, evaluated for the same system size and
boundary conditions as the driven system. This quantity is known exactly
(see, e.g., Refs.~\cite{LSM-61,Katsura-62,Pfeuty-70,BMD-70,BG-85}).
In particular, at the critical point,
$\lim_{L \to \infty} U_0(g_c,L) = 2/\pi$.
As we shall see, this subtraction allows us
to isolate the leading critical scaling behavior.

We also define the time-dependent excess energy
\begin{eqnarray}
  E_{\rm ex}(t)
  &=& E(t)
  - E_0[g,h(t)],
  \label{exenedef}\\
 E(t) & = & 
  {\rm Tr}\big\{\hat \rho(t) \hat H[g,h(t)] \big\},\label{edens}\\  
  E_0(g,h) &=& \langle \Phi_0(g,h) | \hat H(g,h) |
  \Phi_0(g,h) \rangle, \label{eeqist}
\end{eqnarray}
where $E_0(g,h)$ is the ground-state energy associated with the
Hamiltonian $\hat H(g,h)$. Note that when using square-wave periodic
driving protocols, such as Eq.~(\ref{hst}), due to the conservation of the
energy in the  unitary evolution at constant Hamiltonian parameters, the
excess energy turns out to be a step function, which varies only
at multiples of $P/2$.

The dynamic scaling behavior arising from periodic driving of the
longitudinal field in the critical regime will be discussed in
Sec.~\ref{scatheo}.

\subsection{Periodic transverse driving}
\label{traper}

We now consider protocols in which 
$g(t)$ varies periodically at zero longitudinal field $h$. In
this case, the system maintains integrability, i.e., it can be
analyzed exploiting the Jordan-Wigner map into a quadratic fermionic
model~\cite{LSM-61,Katsura-62,Pfeuty-70,BMD-70,BG-85,Sachdev-book},
also known as Kitaev fermionic wire.

Some aspects of the behavior of quantum Ising chains under transverse
periodic driving have already been studied in
Refs.~\cite{BDD-12,RSS-12,LMPPA-17,CZQXX-25,GPJ-26}, also addressing the
possibility of observing dynamic quantum transitions, somehow
unrelated to the underlying equilibrium (ground-state) quantum
transitions of the model.  Here, we instead focus on the scaling
behavior induced by a small-amplitude 
periodic driving of the transverse field across
the continuous quantum transition, i.e., when $g(t)$ oscillates around
$g=g_c=1$, exciting the low-energy critical modes of the underlying
transition.

We assume that $r(t)=g(t)-g_c$ varies periodically
around the transition point $r=0$, at zero
longitudinal field $h=0$ (this is understood in the following
formulas). The dynamic protocol starts at $t=0$ from the 
system ground state $|\Psi(t=0)\rangle=|\Phi_0(g_i)\rangle$
at $g=g_i$.  Then, it evolves unitarily as in the longitudinal case, 
with a periodic
transverse field $g(t)$. For square-wave periodic driving
we define
\begin{equation}
  r(t) \equiv g(t)-g_c = (-1)^{n(t)} B_g, \quad n(t) = \lceil 2\,t/P \rceil,
  \label{hsttr}
\end{equation}
where $B_g > 0$ is the amplitude of the transverse-field
modulation. Single-frequency periodic driving is instead
obtained by taking
\begin{equation}
  r(t) \equiv g(t)-g_c = B_g \cos(\omega t), \quad \omega = \frac{2\pi}{P}.
  \label{hsttrcos}
\end{equation}
In both cases, the evolution starts from the ground state for 
$g_i = g_c + B_g$.
Notice that the longitudinal magnetization remains zero along the
evolution, due to the ${\mathbb Z}_2$ symmetry. Along the evolution
we evaluate the instantaneous expectation value of the
square longitudinal magnetization operator, cf. Eq.~(\ref{mxm2}), or
the transverse magnetization, cf. Eq.~(\ref{mxmz}).

In the following, instead of considering the quantum Ising chain
with a periodic transverse field $g(t)$, we will study
the equivalent Kitaev fermionic wire in the
presence of a periodic chemical potential (see Sec.~\ref{traperkit}).

\section{Dynamic scaling under periodic longitudinal driving}
\label{scatheo}

In this section 
we put forward a dynamic scaling theory for the time dependence of the 
different observables discussed above
under a periodic longitudinal and homogeneous
magnetic field $h(t)$, varying as 
in Eqs.~(\ref{hstcos}) and~(\ref{hst}).
We identify the relevant scaling variables involving the 
amplitude $A$ and the period $P$ of oscillations of the field, 
and determine the relevant power laws.

\subsection{Dynamic scaling theory}
\label{dynscatheo}

We first discuss the scaling theory for finite-size systems,
generalizing the dynamic FSS theory developed for systems in
equilibrium~\cite{FB-72,Barber-83,Privman-90,CPV-14,RV-21} and under 
other, already well-studied, 
out-of-equilibrium conditions, for example after sudden quenches or
along Kibble-Zurek protocols (see, e.g.,
Refs.~\cite{PRV-18,NRV-19,RV-21,TV-22,FRV-22,DV-23,RV-24}).  In the
case of periodic driving $h(t)$ coupled to the order-parameter
operator at a quantum transition, such as the dynamic protocol outlined
in Sec.~\ref{dynscaloper}, a reasonable working hypothesis is that
the amplitude $A$ of the magnetic field $h(t)$ has the same RG
dimension as a static longitudinal field, while the period $P$
scales as a time. The corresponding scaling variables are
therefore $A L^{y_h}$ and $P/L^z$. This conjecture was already
verified numerically in the classical Ising model
driven by a periodic time-dependent magnetic field~\cite{PV-26-per}.

Therefore, to describe the magnetic periodic driving, the dynamic FSS
framework must supplement the standard dynamic FSS variables
\begin{subequations}
  \label{eq:fssvar}
\begin{equation}
  W_r = r L^{y_r},\quad r=g-g_c,\qquad
  W_t = t L^{-z},
  \label{stafssdef}
\end{equation}
with those related to the amplitude and period of the driving, i.e.,
\begin{equation}
    W_A = A L^{y_h},\qquad W_P = P L^{-z}. \label{ufssdef}
\end{equation}
\end{subequations}
Dynamic FSS is obtained in the limit $r\to 0$, $t\to\infty$, $A\to 0$,
$P\to\infty$ and $L\to\infty$ keeping the FSS variables $W_{\#}$
in Eqs.~\eqref{eq:fssvar} fixed.
In the following, we focus on periodic longitudinal drivings at the
critical point $r=0$, thus $W_r=0$. The extension of the dynamic
scaling ansatz to nonzero $W_r$ is simply obtained by adding its
dependence in the scaling functions.

In the dynamic FSS limit at the critical point, the
time-dependent longitudinal magnetization is expected to behave as
\begin{equation}
  M(t,A,P,L)\approx L^{-y_\phi} \widehat{\cal M}_L(W_t,W_A,W_P),
    \label{mtfss}
\end{equation}
where $y_\phi$ is the RG dimension of the order-parameter field,
cf. Eq.~(\ref{yphival}), while the expectation value of its square is
expected to behave as
\begin{equation}
  M_2(t,A,P,L)\approx L^{-2y_\phi} \widehat{\cal C}_L(W_t,W_A,W_P).
  \label{mtfss2}
\end{equation}
The scaling functions ${\widehat{\mathcal M}}_L$ and
${\widehat{\mathcal C}}_L$ are independent, i.e., they are not
trivially related.

The transverse magnetization~\eqref{mxmz} and the
total energy~\eqref{edens} require appropriate subtractions
to isolate their leading critical scaling behavior, as
defined in Eqs.~(\ref{submmxmz}) and (\ref{exenedef}), respectively.
Then, the subtracted transverse magnetization is expected to scale as
\begin{equation}
  U_s(t,A,P,L)\approx  L^{-y_n} \widehat{\cal U}_L(W_t,W_A,W_P),
    \label{mzrfss}
\end{equation}
and the excess energy as
\begin{equation}
  E_{\rm ex}(t,A,P,L)\approx L^{-z} \widehat{\cal E}_L(W_t,W_A,W_P).
    \label{erfss}
\end{equation}

We may also use a different set of scaling variables, which turn
out to be more convenient to describe the thermodynamic limit.
They are defined as
\begin{eqnarray}
  \sigma &=& W_A \, W_P^{{y_h}/{z}}
  = A P^{\kappa}, \quad \kappa = \frac{y_h}{z}=
  \frac{15}{8},
  \label{whdef}\\
  \tau &=& W_t \, W_P^{-1} = t/P. \label{wpdef}
\end{eqnarray}
We can then rewrite the asymptotic dynamic FSS as
\begin{eqnarray}
  M(t,A,P,L)&\approx& A^{y_\phi/y_h} \, {\cal M}_L(W_P,\sigma,\tau),
  \label{mtfssf}\\
  M_2(t,A,P,L)&\approx& A^{2y_\phi/y_h} \, {\cal C}_L(W_P,\sigma,\tau),
  \label{mtfss2f}\\
  U_s(t,A,P,L) &\approx& 
  A^{y_n/y_h} \, {\cal U}_L(W_P,\sigma,\tau),
  \label{mzrfssf}\\
  E_{\rm ex}(t,A,P,L)&\approx& A^{z/y_h} \, {\cal E}_L(W_P,\sigma,\tau),
  \label{erfssf}
\end{eqnarray}
where
\begin{equation}
  \frac{y_\phi}{y_h} = \frac{1}{15},\qquad
  \frac{y_n}{y_h} = \frac{8}{15},\qquad
  \frac{z}{y_h} = \frac{8}{15}.
  \label{listexp}
\end{equation}
These dynamic FSS behaviors at the critical point $g=g_c$ are realized
in the $L\to\infty$ limit keeping $\sigma$, $\tau$, and $W_P$ fixed.

The dynamic FSS functions introduced above are expected to depend on
the type of periodic modulation, for example, whether it is single
frequency or square wave, and also on the type of boundary conditions,
for example, whether they are periodic or open.  On the other hand,
they are expected to be universal with respect to microscopic details
of the model, apart from an overall factor and normalizations of the
arguments. For example, they are expected to be independent of the
parameter $\gamma>0$ appearing in the extension of the quantum Ising
chains to quantum XY chains, see, e.g., Ref.~\cite{RV-21}.

The approach to the asymptotic scaling is expected to be characterized
by power-law suppressed corrections. Under equilibrium conditions with
periodic boundary conditions, scaling corrections to the asymptotic
FSS decay as $L^{-\omega} \sim P^{-\omega}$ where $\omega=2$ is the
leading scaling correction exponent for the 2D Ising universality
class~\cite{PV-02,CHPV-02}. Moreover, one generally expects
$O(L^{-2})$ corrections due to the finite size when using periodic
boundary conditions~\cite{CPV-14}.  Therefore, the leading irrelevant
operators generate $O(L^{-2})$ scaling corrections. Additional
contributions may arise from regular backgrounds and nonlinear
scaling fields, depending on the observable.

Note that the thermodynamic limit of the dynamic scaling, under
periodic longitudinal driving, of the expectation values of $\hat M$,
$\hat M^2$ and $\hat U$ can be straightforwardly obtained from the
dynamic FSS relations written in terms of $\sigma$, $\tau$, and $W_P$,
taking the limit $W_P\to 0$, corresponding to $P\ll L$.  Therefore
such regime, realized by taking $A\to 0$ and $t,P\to\infty$ keeping
$\sigma$ and $\tau$ fixed, entails the asymptotic behaviors
\begin{eqnarray}
  M_\infty(t,A,P)&\approx& A^{{y_\phi}/{y_h}} \, {\cal
    M}_\infty(\sigma,\tau),
  \label{mtthlim} \\
  M_{2,\infty}(t,A,P)&\approx& A^{{2 y_\phi}/{y_h}}
  \, {\cal C}_{\infty}(\sigma,\tau),
  \label{mtthlim2} \\  
  U_{s,\infty}(t,A,P) &\approx& A^{{y_n}/{y_h}}\,
  {\cal U}_{\infty}(\sigma,\tau),
  \label{mtzthlim}
\end{eqnarray}
where the exponents are reported in Eq.~(\ref{listexp}). Moreover, we have 
\begin{equation}
  {\cal M}_\infty(\sigma,\tau)= {\cal M}_L(W_P=0,\sigma,\tau),
  \label{mlminfty}
\end{equation}
and analogous relations for ${\cal C}_\infty$ and ${\cal U}_\infty$.  At fixed
$\sigma > 0$ and finite $\tau$, taking $W_P \to 0$ makes the system
size asymptotically larger than both the initial correlation length
and the distance over which correlations propagate during the
evolution. Moreover, we expect that the excess-energy density
$e_{\rm ex}\equiv E_{\rm ex}/L$ in the thermodynamic limit behaves as
\begin{equation}
  e_{{\rm ex},\infty}(t,A,P)
  \approx A^{{(1+z)}/{y_h}}\, {\cal E}_\infty(\sigma,\tau),
  \label{enethlim}       
\end{equation}
with $(1+z)/y_h = 16/15$.

We finally note that for $t=\tau=0$, the system is in the ground state
of the quantum Ising Hamiltonian for $h=A$ and $g=g_c$, thus, in the
thermodynamic scaling limit, the rescaled observables start from
values that are independent of $\sigma$. In particular, 
the ground-state magnetization behaves as 
\begin{equation}
M_0(g_c,h) = \langle\Phi_0(g_c,h)|
\hat M | \Phi_0(g_c,h) \rangle
\approx {\rm sgn}(h)\, c\,|h|^{y_\phi/y_h}
\label{m0value}
\end{equation}
for $h\to 0$, where $c>0$ is a constant. Therefore, we must have
${\cal M}_\infty(\sigma,\tau=0)=c$, independently of $\sigma$.

\subsection{Numerical results}
\label{numres}

We now support the dynamic FSS behaviors, derived in
Sec.~\ref{dynscatheo}, by numerical results. To this end, we consider
a square-wave periodic driving of $h(t)$, which is more convenient for
numerical simulations, and investigate its effects at the critical
point $g=g_c=1$. We present results for finite chains of up to $L=28$
sites with periodic boundary conditions.

\begin{figure}[!t]
  \includegraphics*[width=0.95\columnwidth]{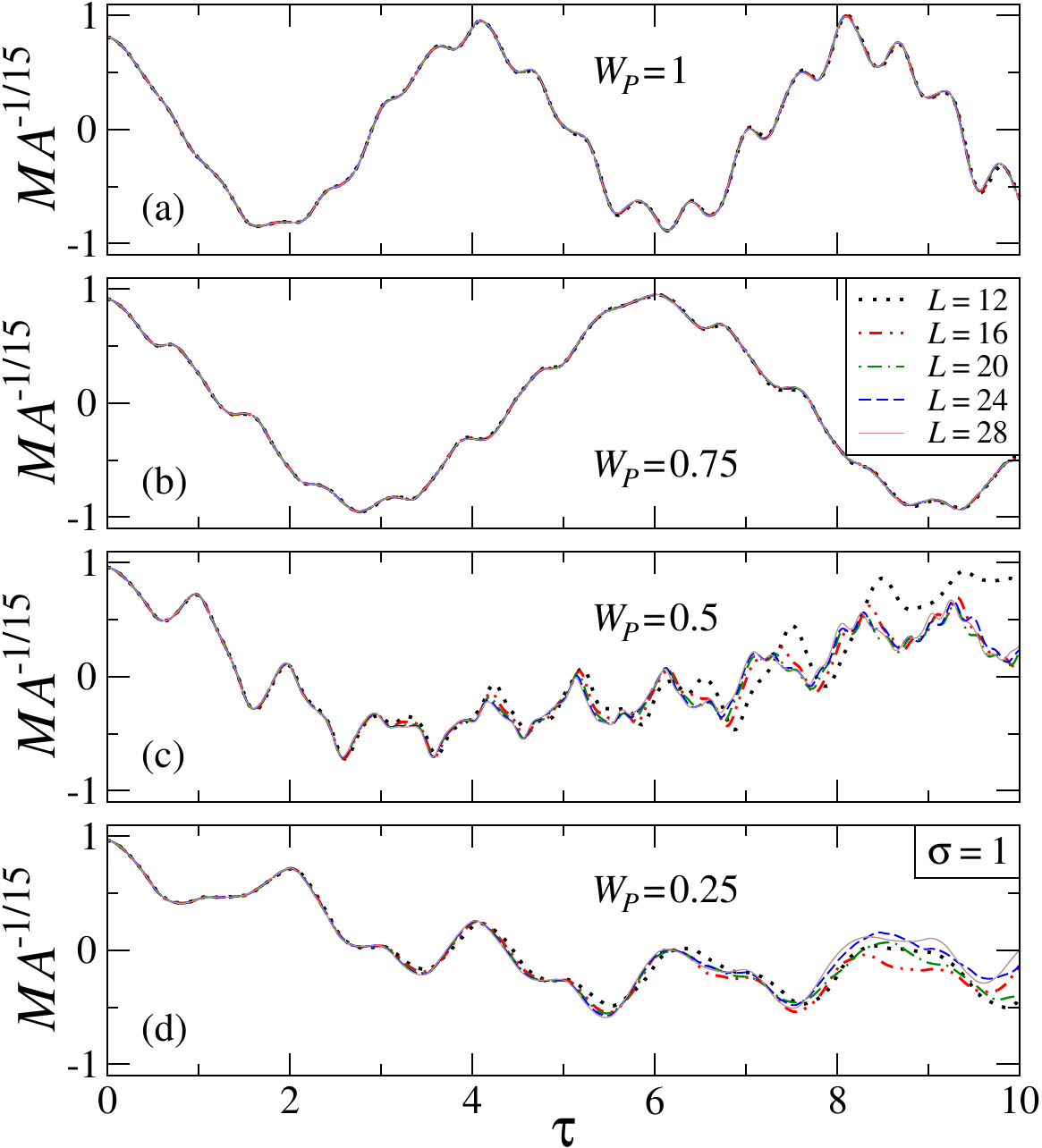}
  \caption{Evolution of the longitudinal magnetization $M$ of a
    quantum Ising chain subject to a periodic square-wave driving of
    the longitudinal field. To check the asymptotic FSS predicted by
    Eq.~(\ref{mtfssf}), we plot $M A^{-1/15}$ vs $\tau\equiv t/P$, at
    fixed $\sigma\equiv A P^{15/8}=1$ and $W_P \equiv P/L= 1$ (a),
    $0.75$ (b), $0.5$ (c), $0.25$ (d). The various curves are for
    different system sizes up to $L=28$ (see legend).}
 \label{loMsig1}
\end{figure}

Calculations are performed in the translation-invariant,
zero-momentum sector, using the symmetry-reduction approach
of Ref.~\cite{PRV-26-qq}. The initial ground state belongs
to this sector, which is preserved by the spatially homogeneous driving.
This restriction substantially reduces the Hilbert-space dimension
without introducing a truncation of the dynamics. The ground state
is obtained by full diagonalization for $L \leq 16$ and by a Davidson
eigensolver for larger systems.
The time-dependent Schr{\"o}dinger equation is integrated using
a fourth-order Runge-Kutta scheme. The time grid is chosen
so that each field switch coincides with a time-step boundary.
Convergence is checked by comparing time steps of $0.001$
and $0.005$ in inverse coupling units, which yield
nearly indistinguishable results for all monitored observables
throughout the investigated time intervals. As an additional
diagnostic, the deviation of the state norm from unity remains
below $\sim 10^{-3}$, including at the largest sizes
and longest times considered.

\begin{figure}[!t]
  \includegraphics*[width=0.95\columnwidth]{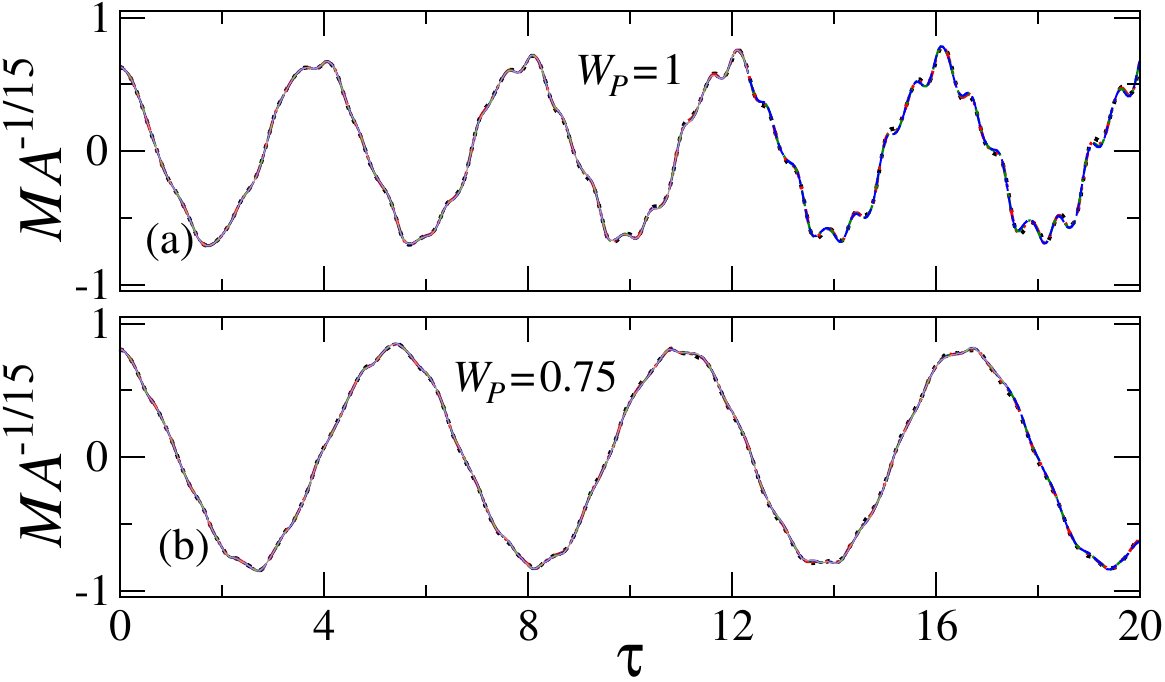}
  \vspace*{2mm} \\
  \includegraphics*[width=0.95\columnwidth]{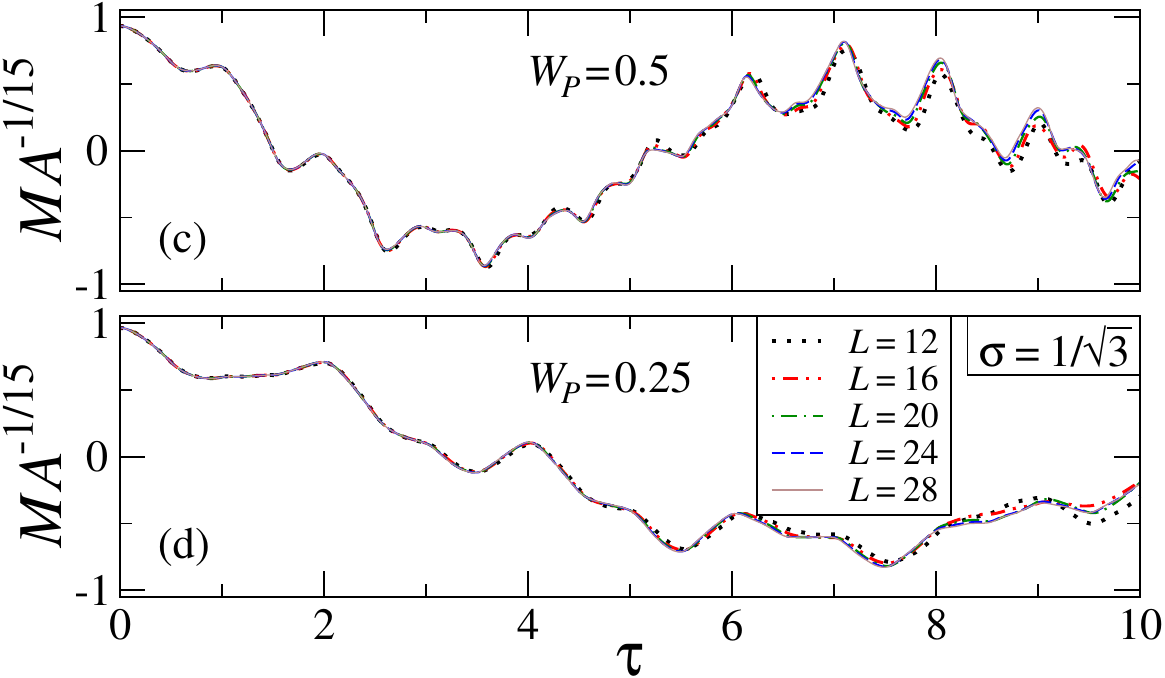}
  \caption{Same as in Fig.~\ref{loMsig1}, but for $\sigma=1/\sqrt{3}$.
    Note that in the two top panels, [for $W_P=1$ (a) and $0.75$ (b)],
    times are running up to $\tau=20$, while in the two bottom panels
    [for $W_P=0.5$ (c) and $0.25$ (d)], times are running
    up to $\tau = 10$.}
  \label{loMsigsq3}
\end{figure}

\begin{figure}[tbp]
  \includegraphics*[width=0.95\columnwidth]{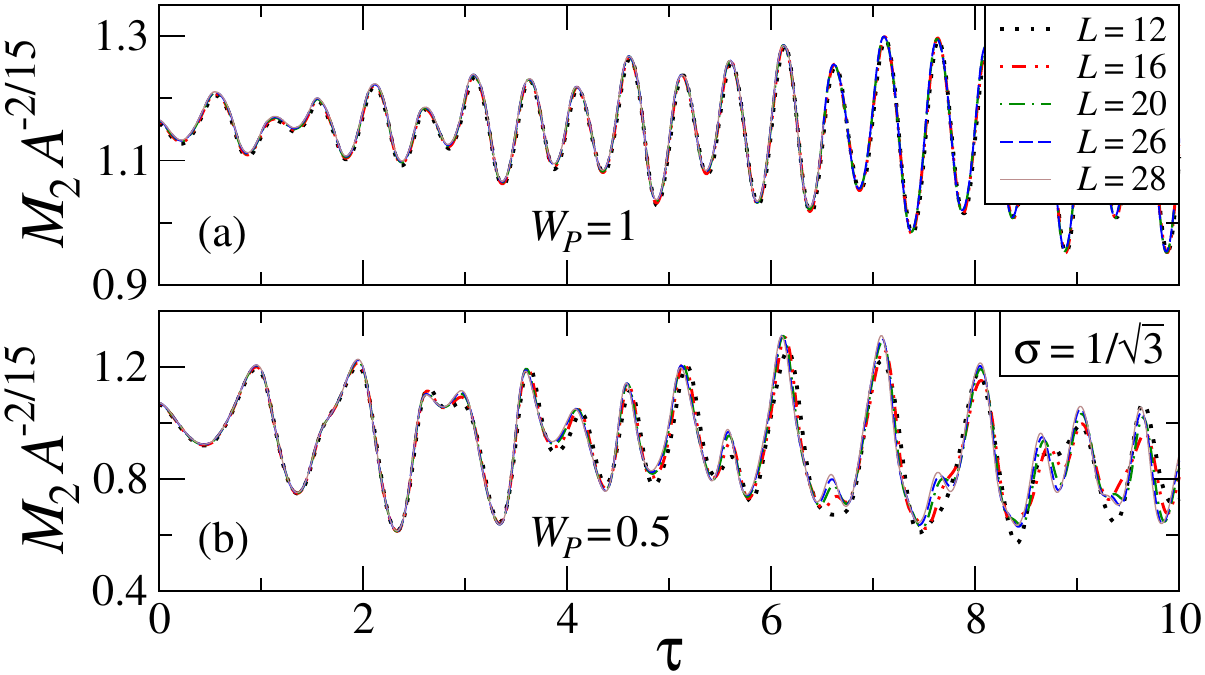}
  \caption{Evolution of the square longitudinal magnetization $M_2$
    under a periodic square-wave driving of the longitudinal field
    $h(t)$. To check the asymptotic FSS predicted by
    Eq.~(\ref{mtfss2f}), we plot $M_2 A^{-2/15}$ vs $\tau$. We fix
    $\sigma=1/\sqrt{3}$ and show results for $W_P = 1$ (a) and $0.5$
    (b).}
 \label{loM2sigsq3}
\end{figure}

\begin{figure}[tbp]
  \includegraphics*[width=0.98\columnwidth]{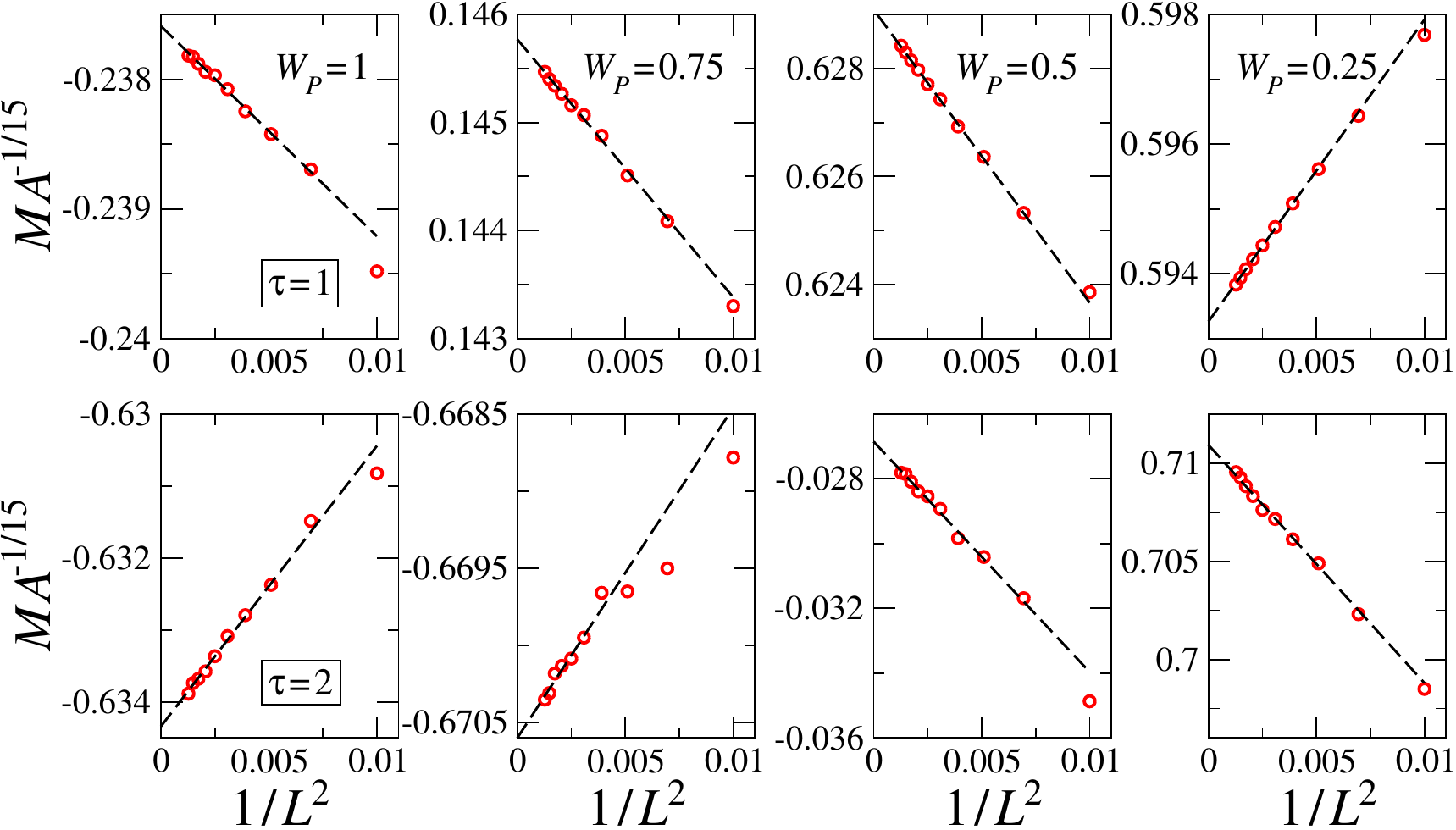}
  \caption{Convergence of the rescaled longitudinal magnetization $M
    A^{-1/15}$ with increasing $L$, plotted against $1/L^{2}$, for two
    values of $\tau = 1$ (top row) and $\tau=2$ (bottom row) and
    $W_P = 1, \, 0.75, \, 0.5, \, 0.25$ (see panels corresponding
    to the various columns).  Data are taken at $\sigma=1/\sqrt{3}$,
    from Fig.~\ref{loMsigsq3}.  Dashed lines show fits to $a+b/L^2$,
    using data with $L \in [14,28]$.  }
 \label{scacorr}
\end{figure}

Figures~\ref{loMsig1}, \ref{loMsigsq3}, and \ref{loM2sigsq3} show the
time evolution of the expectation value $M$ of the longitudinal
magnetization operator and that, $M_2$, of its square operator,
induced by the periodic driving of the longitudinal field $h(t)$, for
some fixed values of $\sigma \equiv A P^{15/8} = 1, \, 1/\sqrt{3}$,
and $W_P\equiv P/L=1,\,0.75,\,0.5,\,0.25$.  They support the dynamic
FSS behavior predicted by Eqs.~(\ref{mtfssf}) and (\ref{mtfss2f}).
Indeed, with increasing lattice size $L$,
the curves of the rescaled longitudinal magnetization $A^{-1/15} M$
and square magnetization $A^{-2/15} M_2$ approach asymptotic scaling
curves when plotted versus the rescaled time $\tau\equiv t/P$,
reproducing the FSS functions ${\cal M}_L(W_P,\sigma,\tau)$ and
${\cal C}_L(W_P,\sigma,\tau)$ respectively, at fixed values
of $\sigma$ and $W_P$.
Note that already the initial rescaled magnetization depends on $\sigma$
at fixed $W_P$ [compare Figs.~\ref{loMsig1}(a)
  with~\ref{loMsigsq3}(a)]. This is consistent with the discussion
around Eq.~\eqref{m0value}: independence of $\sigma$ is recovered
only in the thermodynamic scaling limit $W_P \to 0$ at fixed $\sigma$.

Scaling corrections appear to generally increase with increasing
$\tau$, and their size also depends on $W_P$, requiring larger and
larger lattice sizes to observe convergence, suggesting that the
convergence is not uniform in $\tau$. In particular, the available
data up to $L=28$ allow us to achieve clear evidence of scaling up to
$\tau \approx 20$ for $W_P=1,\,0.75$, and up to $\tau\approx 10$ for
$W_P=0.5,\,0.25$.  Figure~\ref{scacorr} shows that the size-dependence of
the rescaled magnetization is consistent with leading $O(L^{-2})$ 
corrections over the available size range, as expected from general
arguments (see Sec.~\ref{dynscatheo}).

We note that the scaling function ${\cal M}_L(W_P,\sigma,\tau)$ shows
an approximate periodicity with respect to the scaling variable
$\tau$, which is not directly related to the period $P$ of the
driving, corresponding to $\tau=1$, and it is apparently strongly
dependent on the scaling variables $\sigma$ and $W_P$. The scaling
function ${\cal C}_L(W_P,\sigma,\tau)$ associated with $M_2$ appears
even less regular. Therefore, the numerical data do not show evidence
of synchronization. However, this is not unexpected: already the
single spin model with periodic external fields does not generally
display synchronization (see, e.g., Ref.~\cite{GH-98}).

\begin{figure}[!t]
\includegraphics*[width=0.95\columnwidth]{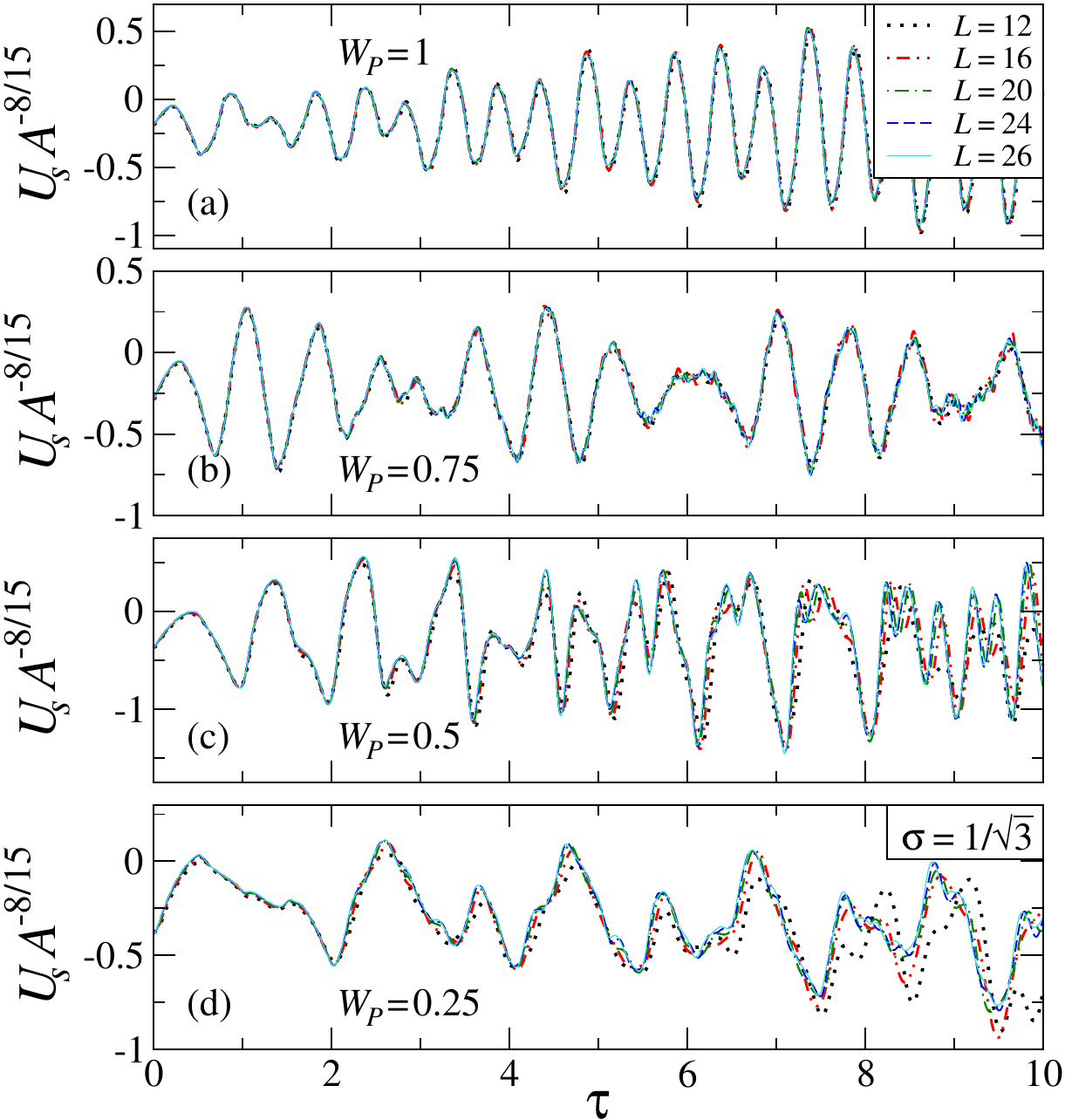}
  \caption{Evolution of the subtracted transverse magnetization $U_s$
    under the periodic square-wave driving $h(t)$.  To verify the FSS
    predicted by Eq.~(\ref{mzrfssf}), we plot $U_s A^{-8/15}$ vs
    $\tau$.  We fix $\sigma=1/\sqrt{3}$ and show results for $W_P = 1$
    (a), $0.75$ (b), $0.5$ (c), $0.25$ (d).}
 \label{loUssigsq3}
\end{figure}

\begin{figure}[!t]
  \includegraphics*[width=0.95\columnwidth]{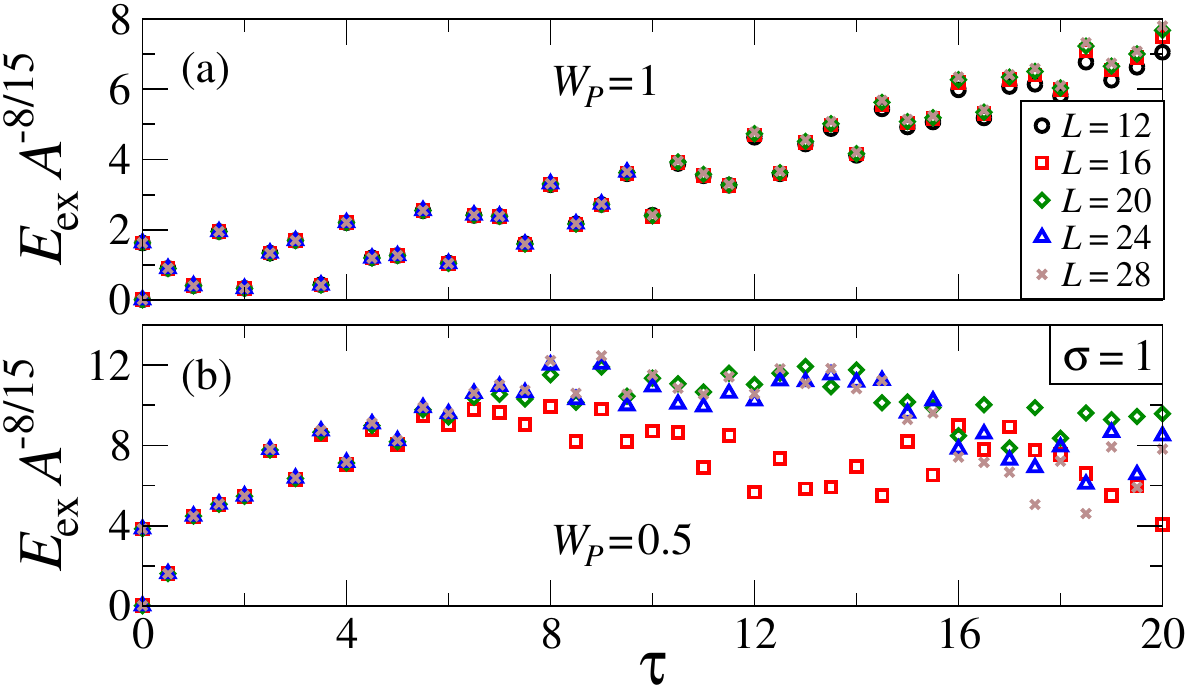}
  \caption{Evolution of the excess energy $E_{\rm ex}$ under the
    periodic square-wave driving $h(t)$. $E_{\rm ex}$ is evaluated
    stroboscopically at times $t=(n/2)P$, with $n = 1, \, 2, \,
    \ldots$, and data are sampled immediately after each field
    switch. To verify the FSS predicted by Eq.~(\ref{erfssf}), we plot
    $E_{\rm ex} A^{-8/15}$ vs $\tau$.  We fix $\sigma=1$ and show
    results for $W_P = 1$ (a) and $0.5$ (b).}
 \label{loEexsig1}
\end{figure}

Clear evidence of dynamic FSS is also observed for the subtracted
quantities $U_s$ and $E_{\rm ex}$, defined in Eqs.~\eqref{submmxmz}
and~\eqref{exenedef} respectively (see Figs.~\ref{loUssigsq3} and
\ref{loEexsig1}), whose time dependence nicely confirms the scaling
behaviors (\ref{mzrfssf}) and (\ref{erfssf}), at least for not too
large rescaled times, analogously to the case of the longitudinal
magnetization.
For the excess energy, the data are sampled immediately after
each field switch, at $t_n = n P/2$ with $n = 1,2,\ldots$,
using the post-switch Hamiltonian and its ground-state energy.

In conclusion, the numerical results presented here support the
dynamic FSS theory for periodic longitudinal drivings developed in
Sec.~\ref{dynscatheo}, at least for moderately large scaling times
(say, $\tau\lesssim 20$). However, we expect dynamic FSS to hold at
larger scaling times, as well. As already noted, evidence for
scaling becomes weaker as $\tau$ increases, with increasingly
significant scaling corrections. This suggests that progressively
larger system sizes are required to observe scaling at larger values
of $\tau$. Unfortunately, computational limitations restrict
our analysis to chains with $L \leq 28$, preventing us from
investigating the asymptotic large-$\tau$ behavior of the scaling
functions.

The available system sizes are insufficient to reliably investigate
the thermodynamic limit and test the scaling predictions
in Eqs.~\eqref{mtthlim}, \eqref{mtthlim2}, \eqref{mtzthlim},
and~\eqref{enethlim}. Nevertheless, we expect the predicted
scaling behaviors of the longitudinal magnetization,
its second moment, the subtracted transverse magnetization,
and the excess-energy density to hold, as they follow naturally
from the thermodynamic limit of the dynamic FSS framework.

We mention that other numerical methods, such as tensor-network
techniques, can be used to investigate periodically driven dynamics
directly in the thermodynamic limit. These include the infinite
time-evolving block decimation (iTEBD) algorithm~\cite{Vidal-07} and
the time-dependent variational principle (TDVP) applied to uniform
matrix product states~\cite{HCOPVV-11}.  Tensor-network methods have
also been successfully employed to investigate relaxation and
thermalization in infinite nonintegrable Ising chains~\cite{BCH-11}.
However, verifying scaling behaviors and subtle dynamical features
over many driving periods, corresponding here to times of several
hundred inverse coupling units, requires stringent numerical
accuracy. Entanglement growth can make this challenging. Working
directly in the thermodynamic limit would eliminate finite-size
effects, but would require careful convergence checks with respect to
both bond dimension and time-evolution accuracy.  Our finite-size
approach~\cite{PRV-26-qq} is complementary to these methods: it avoids
truncations of the many-body Hilbert space within the relevant
symmetry sector and, with independently controlled time-evolution
errors, provides a reliable benchmark for the delicate dynamical
features considered here.

\section{Fermionic wires with periodic chemical potential}
\label{traperkit}

\subsection{The Kitaev fermionic wire}
\label{Kitmod}

By means of a nonlocal Jordan-Wigner
transformation~\cite{LSM-61,Katsura-62,Pfeuty-70,BMD-70,BG-85}, the quantum
Ising chain with $h=0$ can be exactly mapped onto the Kitaev fermionic
wire
\begin{equation}
  \hat H(\mu) = - J \sum_{x} \big( \hat c_x^\dagger \hat
  c_{x+1}^{\phantom\dagger} + \hat c_x^\dagger \hat
  c_{x+1}^\dagger+{\rm h.c.}  \big) - \mu \hat{N},
  \label{kitaev}
\end{equation}
where $\hat c_x$ is the fermionic annihilation operator associated
with the $x$-th site of the chain, and 
\begin{equation}
  \hat{N} = \sum_x \hat{n}_x, \qquad \hat n_x=\hat c_x^\dagger \hat
  c_x^{\phantom\dagger},
  \label{partnumb}
\end{equation}
is the particle-number operator. The Hamiltonian parameter $\mu$
is the chemical potential and $J$ sets the energy scale.
Hereafter, we set $J=1$.  We also report the relations 
\begin{equation}
  g=-\mu/2, \quad \hat\sigma^{(3)}_x = 1 - 2 \hat{n}_x,
  \label{sign}
\end{equation}
between the two models.  Therefore, the Kitaev fermionic wire
undergoes a continuous quantum transition at $\mu = \mu_c = - 2$,
belonging to the 2D Ising universality class.
Deviations from the critical point are parametrized by $w = \mu - \mu_c$,
which is related to the Ising parameter $r = g - g_c$ by $r = -w/2$.
The RG dimension of the
fermionic operators $\hat c_{x}$ is $y_c = 1/2$, and that of the
particle density $\hat n_x$ is $y_n=1$.

Although the bulk behavior of the quantum Ising chain and of the 
Kitaev fermionic wire are analogous in the infinite-volume limit (and so are
their phase diagrams), some finite-size features differ
significantly.  For example, the nonlocality of the Jordan-Wigner
transformation implies that the Ising chain with periodic
or antiperiodic boundary conditions is not mapped onto the Kitaev wire with
analogous boundary conditions and that 
the correspondence between the two models also depends on the parity of the
particle-number eigenvalue (see, e.g., Refs.~\cite{LSM-61,Katsura-62,
  Pfeuty-70, CPV-14}). In contrast, these problems disappear
when considering open boundary conditions for both of them.

In our FSS analysis, we consider Kitaev chains of even length $L$
with antiperiodic boundary conditions,
$\hat c^{(\dagger)}_{x+L} = -\hat c^{(\dagger)}_x$.
The allowed momenta are therefore $k = (2m+1) \pi /L$, with
$m = 0,\ldots, L-1$, so the smallest positive momentum
is $k_m=\pi/L$.
These boundary conditions allow a convenient momentum-space
description in terms of paired modes with opposite momenta.

\subsection{The periodic driving protocol}
\label{Kitprot}

We now consider periodic modulations of the chemical potential
around $\mu_c = -2$. We denote their amplitude with $B > 0$, which is 
the analogue of the transverse-field amplitude $B_g$ introduced
in Sec.~\ref{traper}.
Under the mapping~\eqref{sign}, the amplitudes satisfy $B = 2 B_g$,
while deviations from criticality have opposite signs.
We consider a single-frequency modulation 
\begin{equation}
  w(t) = \mu(t) - \mu_c = B \cos(\omega t), \quad \omega = 2\pi/P,
  \label{rtmucos}
\end{equation}
and a  discontinuous square-wave modulation 
\begin{equation}
  w(t) = \mu(t) - \mu_c = (-1)^{n(t)} B, \quad n(t) = \lceil 2 t/P \rceil.
  \label{rtmu}
\end{equation}
In both cases, we start from the ground state at $\mu_i = \mu_c + B$.
The initial ground state has even fermion parity, which is conserved
throughout the evolution.
Under the mapping~\eqref{sign}, this corresponds to an initial
transverse field $g_i = g_c - B_g$, with $B_g = B/2$. Thus, the 
initial configuration considered here has $r = g_i - g_c < 0$, while 
in Sec.~\ref{traper} we considered a state with $r > 0$. This 
does not affect the analysis of the scaling behavior.
For square-wave driving, we use the same measurement convention
as in Sec.~\ref{dynscaloper}: values reported at switching times refer
to the right-hand limit, with the post-switch chemical potential.

To monitor the spatial correlations arising from the periodic driving,
we consider the fermionic correlation functions
\begin{eqnarray}
G_e(x-y,t) & \!\! = \!\! & {\rm Tr}[\hat\rho(t)\, (\hat c_x^\dagger \hat c_{y} 
+ \hat c_{y}^\dagger \hat c_{x})], \label{gtf} \\
  G_o(x-y,t) & \!\! = \!\! & {\rm Tr}[\hat\rho(t)\,(
    \hat c_{x} \hat c_{y}
    +  \hat c_y^\dagger \hat c_{x}^\dagger)].
  \label{ptf}
\end{eqnarray}
Here site indices are extended using the antiperiodic convention
$\hat c^{(\dagger)}_{x+L} = -\hat c^{(\dagger)}_x$. Translation symmetry,
with this boundary twist understood, makes the correlators functions
of the separation alone. Their extension to separations outside
the chain satisfies $G_{e/o}(x+L,t) = -G_{e/o}(x,t)$.
One can easily check
that $G_o(x,t)$ is an odd function of $x$, while $G_e(x,t)$ is
even. We also define related low-momentum quantities that can be
useful for the numerical analysis, such as
\begin{eqnarray}
  Q_e(t) = \sum_x G_e(x,t) \cos(\pi x/L),
  \label{qedef} \\  
  Q_o(t) = \sum_x G_o(x,t) \sin(\pi x/L),
  \label{qodef}
\end{eqnarray}
where the sums run over one complete set of signed separations,
chosen as $x = -L/2, \ldots ,L/2-1$. The sine and cosine weights
correspond to the smallest positive momentum allowed by the
antiperiodic boundary conditions, $k_m = \pi/L$, and reflect
the odd and even spatial parity of $G_o$ and $G_e$, respectively.
These quantities admit a simple momentum-space interpretation.
Using the Fourier transform
$\hat c_x=L^{-1/2}\sum_k e^{ikx}\hat c_k$,
and defining
\begin{equation}
  n_k(t)={\rm Tr}[ \hat \rho(t) \, \hat c_k^\dagger\hat c_k ], \qquad
  f_k(t) = {\rm Tr}[ \hat \rho(t) \, \hat c_k\hat c_{-k} ],
\end{equation}
we obtain
\begin{equation}
  Q_e(t) = 2 \, n_{k_m}(t),
  \qquad
  Q_o(t)=-2\, \mathrm{Im}\, f_{k_m}(t),
\end{equation}
where we used $n_k(t)=n_{-k}(t)$. Thus, $Q_e$ and $Q_o$ probe the occupation
and anomalous coherence of the lowest-momentum pair, respectively.
In contrast, the particle density would involve all momentum modes:
$n_x(t) = {\rm Tr} [\hat \rho(t) \, \hat n_x] = L^{-1}\sum_k n_k(t)$.

When the system size is much larger than the spatial scales
contributing appreciably to the correlations, the sine weight in $Q_o$
can be expanded, yielding
\begin{eqnarray}
  Q_o(t) \approx  \frac{\pi}{L} \sum_x x \, G_o(x,t), \label{q1def2}
\end{eqnarray}  
related to the first moment of the fermionic correlation $G_o$.
Therefore, the quantity that is expected to have a nontrivial
thermodynamic limit is
\begin{equation}
  W(t) = L \, Q_o(t).
  \label{wtdef}
\end{equation}

One may also monitor the dynamics using the expectation values of the
particle-density operator and instantaneous Hamiltonian of the system.
As we shall see, to observe dynamic scaling under the periodic driving,
we need to consider subtracted quantities, such as
the subtracted particle density
\begin{eqnarray}
  D(t) = {\rm Tr}\big[\hat \rho(t) \hat n_x\big] -
  \langle \Phi_0[\mu(t)] | \hat n_x | \Phi_0[\mu(t)] \rangle,
  \label{pdmzz}
\end{eqnarray}
where we used the translation invariance of the system with
antiperiodic boundary conditions.
We also consider the excess energy $E_{\rm ex}$, defined as
\begin{eqnarray}
  E_{\rm ex}(t) &=& {\rm Tr}\big\{\hat \rho(t) \hat H[\mu(t)] \big\}
   - E_0[\mu(t)] = L \,e_{\rm ex}(t),\quad
  \label{hhmzzal}\\
  E_0(\mu) &=&
  \langle \Phi_0(\mu) | \hat H(\mu) | \Phi_0(\mu) \rangle ,
\label{eeqkit}
\end{eqnarray}
where $e_{\rm ex}$ is the excess-energy density. 

For smooth driving, the subtracted particle density $D(t)$ and
the excess-energy density $e_{\rm ex}(t)$ satisfy an exact relation.
The time dependence of the Kitaev Hamiltonian~\eqref{kitaev}
enters only through the chemical potential, so that
$\partial_t \hat H=-\dot{\mu}(t)\hat N$.  Unitary evolution and
the Hellmann–Feynman theorem then yield, respectively,
\begin{subequations}
\begin{eqnarray}
  && \frac{d}{dt}\,{\rm Tr}\big\{\hat\rho(t)\hat H[\mu(t)]\big\}
  = -\dot{\mu}(t)\,{\rm Tr}\big[\hat\rho(t)\hat N\big], \\
  &&\frac{d}{dt}\,E_0[\mu(t)] = -\dot{\mu}(t)\,
  \langle\Phi_0[\mu(t)]|\hat N|\Phi_0[\mu(t)]\rangle . \qquad
\end{eqnarray}
\end{subequations}
Subtracting these expressions, dividing by $L$, and using
translation invariance, we obtain
\begin{equation}
  \frac{d e_{\rm ex}}{dt} = -\dot{\mu}(t)\, D(t) .
  \label{eex_d-rel}
\end{equation}
This identity holds also at finite $L$. Note that, for square-wave
driving, the excess energy is constant between switches and can
change discontinuously at each switch.

\subsection{Dynamic finite-size scaling theory}
\label{Kitdynfsssca}

Following the dynamic scaling framework outlined in
Sec.~\ref{scatheo}, we introduce the scaling variables
\begin{equation}
  X = x/L,\;
  W_B = B L^{y_r}, \; W_P = P/L^{z},\; W_t = t/L^z,
  \label{kitfssdef}
\end{equation}
where $y_r=1/\nu=1$ and $z=1$.  Extending the dynamic FSS framework
developed for nonequilibrium critical behavior (see, e.g.,
Refs.~\cite{PRV-18,RV-21,TV-22,FRV-22,DV-23,RV-24}),
we expect the fermionic correlation functions to scale as
\begin{eqnarray}
  G_{e/o}(x,t,B,P,L)\approx L^{-2y_c} \widehat{\cal G}_{e/o,L}(X,W_t,W_B,W_P),
  \label{g12fssz}
\end{eqnarray}
with $2y_c=1$. This scaling form is valid at fixed nonzero values
of $X=x/L$, but not directly for $X = 0$, where
the contact value $G_e(0,t)=2 \, n_x(t)$ is dominated by the 
regular contribution of order one. For the low-momentum quantities
defined above, we expect
\begin{eqnarray}
  Q_{e/o}(t,B,P,L)&\approx& L^{1-2y_c}\widehat{\cal Q}_{e/o,L}(W_t,W_B,W_P)
  \nonumber
  \\ &\approx &
  \widehat{\cal Q}_{e/o,L}(W_t,W_B,W_P),
  \label{chifssz}
\end{eqnarray}
where we used the fact that the exponent of the prefactor $L^{1-2
  y_c}$ vanishes.  The above dynamic FSS is expected to be realized in
the limit $t\to\infty$, $B\to 0$, $P\to\infty$ and $L\to\infty$
keeping the FSS variables $W_B$, $W_P$, and $W_t$ fixed.

We may also use a different set of scaling variables, which turn
out to be more convenient to describe the thermodynamic limit.
They are $\tau=t/P$ and 
\begin{equation}
  \varrho = W_B \, W_P^{y_r/z} = B \, P,
  \qquad \upsilon = W_B^{1/y_r}\,X =  B\, x.
  \label{kitscadef}
\end{equation}
They allow us to write the FSS behaviors in the equivalent way
\begin{eqnarray}
  G_{e/o}(x,t,B,P,L)&\approx& B \,{\cal
    G}_{{e/o},L}(W_P,\upsilon,\varrho,\tau),
  \label{geofssz2}\\
  Q_{e/o}(t,B,P,L)&\approx&  {\cal Q}_{e/o,L}(W_P,\varrho,\tau).
  \label{chifssz2}
\end{eqnarray}
The dynamic FSS of the particle density is more complicated, since
it requires the subtraction of the instantaneous ground-state value,
as defined in Eq.~(\ref{pdmzz}) (see, e.g.,
Refs.~\cite{RV-24,PV-26-esub}). Such subtraction allows us to
disentangle the scaling part of the particle density from 
the background analytic contribution,
as already observed for
other dynamic protocols, for instance for quench and Kibble-Zurek
protocols~\cite{TV-22,RV-24,PV-26-esub}.  Its dynamic FSS behavior can be
derived as for the fermionic correlations, leading to
\begin{eqnarray}
  D(t,B,P,L)&\approx& L^{-y_n} \widehat{\cal D}_L(W_t,W_B,W_P)\nonumber\\
  &\approx& B^{{y_n}/{y_r}} {\cal D}_L(W_P,\varrho,\tau).
  \label{pdrfssz2}
\end{eqnarray}
For square-wave driving, this scaling ansatz needs to be clarified
at switching times. While the density of the evolving state remains
continuous, the instantaneous ground-state subtraction changes
discontinuously. Writing $n_0(\mu,L)$ for the equilibrium particle density,
the jump is exactly
\begin{equation}
  D(t_s^+) - D(t_s^-) = n_0(\mu_-,L) - n_0(\mu_+,L),
  \label{D-jump}
\end{equation}
where $\mu_-$ and $\mu_+$ denote the chemical potentials before
and after the switch. Near criticality, this equilibrium difference
contains a logarithmic contribution~\cite{RV-24}, so that the magnitude
of the jump of $D/B$ grows logarithmically with $L$ in the FSS limit.
Thus, Eq.~\eqref{pdrfssz2} should not be interpreted as a uniform
scaling description across the switching times.

For the excess energy defined in Eq.~\eqref{hhmzzal}, analogous
scaling arguments suggest the ansatz
\begin{eqnarray}
  E_{\rm ex}(t,B,P,L)&\approx& L^{-z} \widehat{\cal E}_L(W_t,W_B,W_P)
  \nonumber\\
  &\approx& B^{{z}/{y_r}} {\cal E}_L(W_P,\varrho,\tau).
  \label{kierfssz}
\end{eqnarray}
As shown below, this ansatz is supported by the numerical results
for smooth single-frequency driving. In this case, the exact
density–energy relation~\eqref{eex_d-rel} takes the rescaled form
\begin{equation}
  \frac{d}{d\tau} \bigg(\frac{e_{\rm ex}}{B^2} \bigg)
  = 2 \pi \sin(2 \pi \tau) \frac{D}{B} \,.
  \label{eex-D}
\end{equation}
At fixed $W_B = B\,L$, the relation $e_{\rm ex}/B^2 = (E_{\rm ex}/B)/W_B$
shows that this identity also constrains the finite-size scaling
functions. Since the initial excess energy vanishes, integration
of Eq.~\eqref{eex-D} determines the energy scaling function
from the particle-density scaling function.

For square-wave driving, however, the scaling ansatz~\eqref{kierfssz}
requires a clarification. As shown in App.~\ref{endensque}, the initial
sudden quench generates a logarithmic contribution to the excess energy
that is not removed by the instantaneous ground-state subtraction.
This contribution persists throughout the first half-period,
when the Hamiltonian is constant, thus preventing the pure
power-law scaling form~\eqref{kierfssz} from holding over that interval.

\subsection{Dynamic scaling in the thermodynamic limit}
\label{Kitdynsca}

The thermodynamic limit of the fermionic correlation functions
$G_{e/o}$ can be formally obtained by taking the limit $W_P\to 0$ in
Eq.~(\ref{geofssz2}) keeping $\varrho$, $\tau$ and $\upsilon$ fixed,
obtaining
\begin{equation}
  G_{e/o,\infty}(x,t,B,P) \approx
  B \,{\cal G}_{{e/o},\infty}(\upsilon,\varrho,\tau).
  \label{g12thlim}
\end{equation}
Therefore, dynamic scaling is expected to be observed in the limit
$B\to 0$ keeping $\upsilon,\varrho,\tau$ constant.  As already
mentioned in Sec.~\ref{Kitprot}, we expect that $W(t) = L Q_o(t)$,
cf.~Eq.~(\ref{wtdef}), has a nontrivial thermodynamic limit.  Its
scaling behavior can be easily derived from the FSS of the related
quantity $Q_o$, obtaining
\begin{eqnarray}
  W_\infty(t,B,P)  \approx B^{-1} {\cal W}_\infty(\varrho,\tau).
  \label{wtsca}
\end{eqnarray}
The subtracted particle density in the thermodynamic limit is expected
to develop the dynamic scaling
\begin{eqnarray}
  D_\infty(t,B,P) \approx B \, {\cal D}_{\infty}(\varrho,\tau).
  \label{pdzthlimzz}
\end{eqnarray}
For square-wave driving, this scaling form is expected to hold at fixed
rescaled times away from the switches; convergence need not be
uniform in their vicinity.
Moreover, in the case of single-frequency driving, the excess-energy
density $e_{\rm ex}$ of Eq.~\eqref{hhmzzal}, in the thermodynamic
limit, is expected to show the dynamic scaling
\begin{equation}
  e_{{\rm ex},\infty}(t,B,P) 
  \approx B^2 \, {\cal E}_\infty(\varrho,\tau).
  \label{enethlimkit}       
\end{equation}
For cosine driving, Eq.~\eqref{eex-D} also relates the density
and excess-energy scaling functions in the thermodynamic limit.
For square-wave driving, the logarithmic contribution discussed
in App.~\ref{endensque} prevents the excess-energy density
from obeying the pure power-law scaling form in Eq.~\eqref{enethlimkit},
analogously to the violation of the FSS ansatz in Eq.~\eqref{kierfssz}.

We finally remark that convergence to the thermodynamic limit is
expected to be nonuniform in time: progressively larger system sizes
are required to approach this limit as time increases, and hence as
the rescaled time $\tau$ increases. Indeed, the propagation of
information across a spatial interval of length $\ell$ is expected to
take a time $t\sim \ell^z$, with $z=1$. Consequently, finite-size
effects are expected to become eventually relevant for $t\gtrsim L$.
Therefore, the thermodynamic limit in the dynamic FSS relations is
expected to be effectively realized when $t/L=\tau W_P\ll 1$.
In addition, the system size must be much larger than the initial
correlation length, requiring $BL \gg 1$. At fixed $\varrho$,
both conditions are satisfied as $W_P \to 0$ at finite $\tau$.

\subsection{Numerical results}
\label{numkitaev}

We now provide evidence that numerical computations support the above
dynamic scaling behaviors of the Kitaev fermionic wires under periodic
driving of their chemical potential, for both single-frequency
and square-wave cases. The quadratic structure of the
Hamiltonian~\eqref{kitaev} allows the dynamics to be decomposed
into independent sectors associated with pairs of opposite momenta.
Starting from the ground state,
we evolve the corresponding two-component states and
reconstruct the observables from their momentum-space contributions.
The allowed momenta are fixed by the antiperiodic boundary conditions.
This approach allows us to study chains with up to $L = 204800$ sites
and investigate scaling in the thermodynamic limit.

\begin{figure}[tbp]
  \includegraphics[width=0.95\columnwidth]{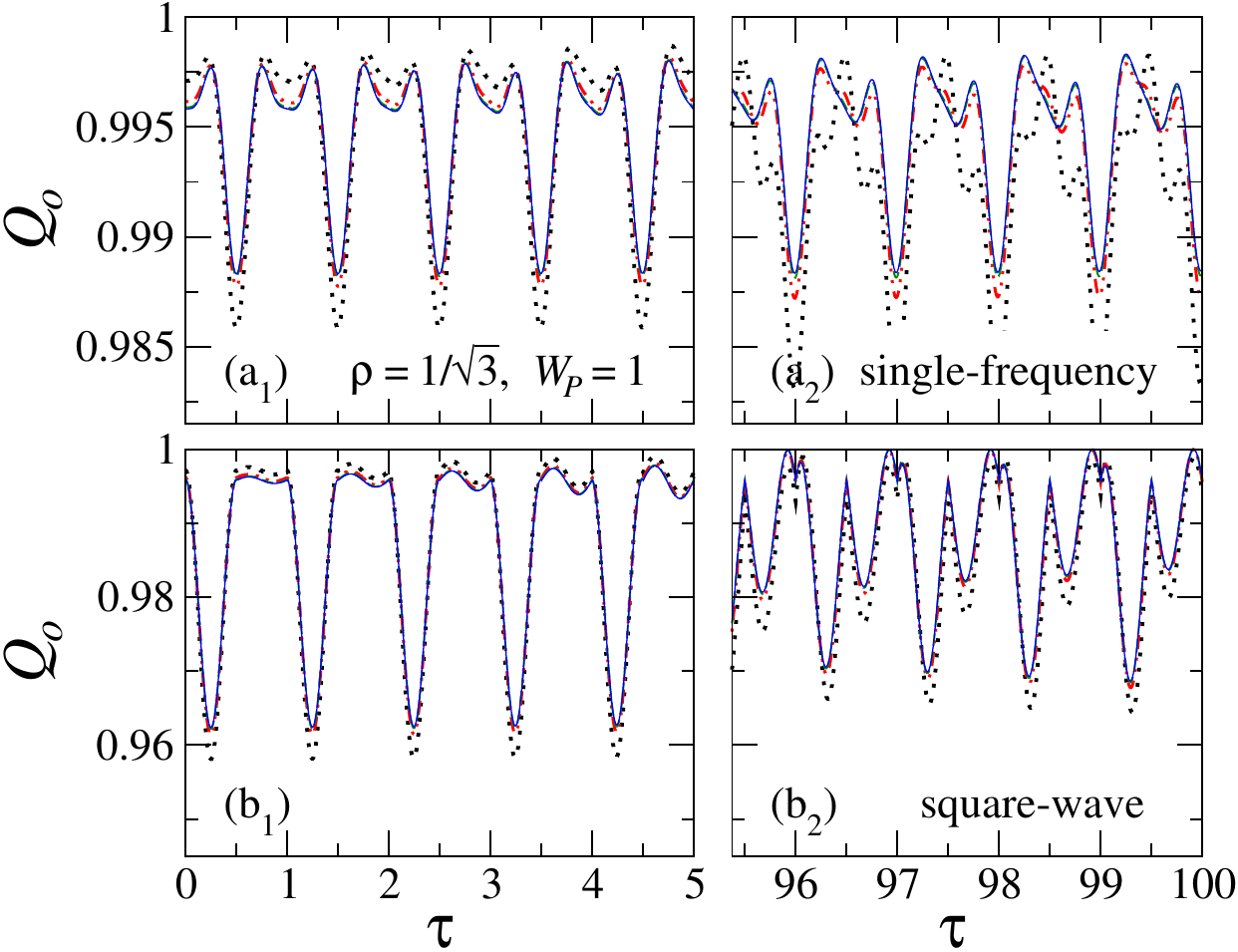}
  \vspace*{3mm} \\
  \includegraphics[width=0.95\columnwidth]{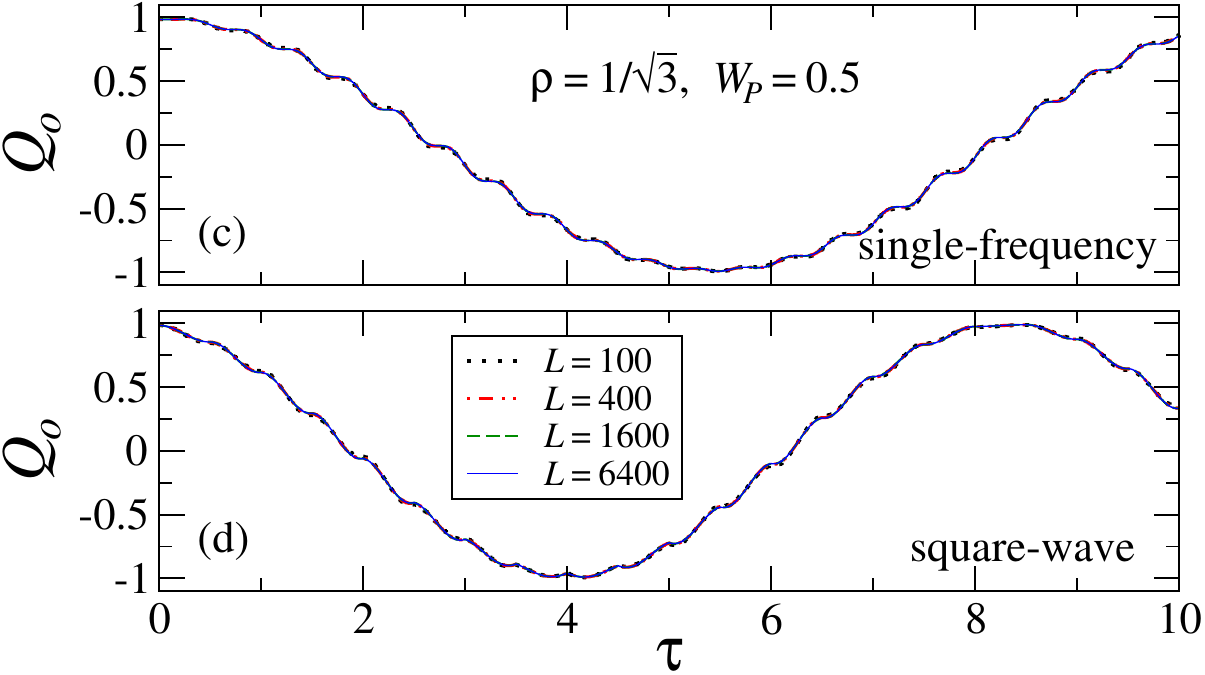}
  \caption{The quantity $Q_o$ defined in Eq.~(\ref{qodef}) vs $\tau=t/P$
    of a Kitaev fermionic chain under single-frequency
    [(${\rm a}_1$), (${\rm a}_2$), and (c)] and square-wave
    [(${\rm b}_1$), (${\rm b}_2$),
      and (d)] driving of the chemical potential, for
    $\varrho=1/\sqrt{3}$ and $W_P=1$ [(${\rm a}_1$), (${\rm a}_2$),
      (${\rm b}_1$), (${\rm b}_2$)] or $0.5$ [(c)-(d)].
    The various curves, for different system sizes (see legend),
    appear to approach a single scaling curve supporting the dynamic FSS
    in Eq.~(\ref{chifssz2}).
    Note that panels (${\rm a_2}$) and (${\rm b_2}$) focus on data
    at long times, up to $\tau=100$.}
  \label{Qofss}
\end{figure}

For both driving protocols, the time evolution is computed
using a fourth-order Runge–Kutta scheme. For square-wave driving,
each time step lies entirely within a constant-field interval,
with field switches aligned with time-step boundaries.
For cosine driving, the Hamiltonian is evaluated at the initial,
intermediate, and final times required by the Runge–Kutta stages.
Time-step convergence is checked by comparing steps of $0.001$ and $0.005$
in inverse coupling units, yielding nearly indistinguishable results
for all monitored observables over the investigated time intervals.

For cosine driving, the excess energy is evaluated at the same
sampling times as the other observables, using the instantaneous
Hamiltonian and its ground-state energy. For square-wave driving,
the excess energy is sampled immediately after each field switch.
Whenever a density measurement coincides with a switching time,
the equilibrium density subtracted in $D$ is likewise evaluated
at the post-switch chemical potential.

\subsubsection{Dynamic finite-size scaling}
\label{kitfss}

We first show numerical results confirming the dynamic FSS behaviors
put forward in Sec.~\ref{Kitdynfsssca} for both single-frequency and
square-wave drivings. They are obtained in the large-$L$ limit keeping
the scaling variables $\varrho$, $\tau$, and $W_P$ constant.

\begin{figure}[tbp]
  \includegraphics[width=0.95\columnwidth]{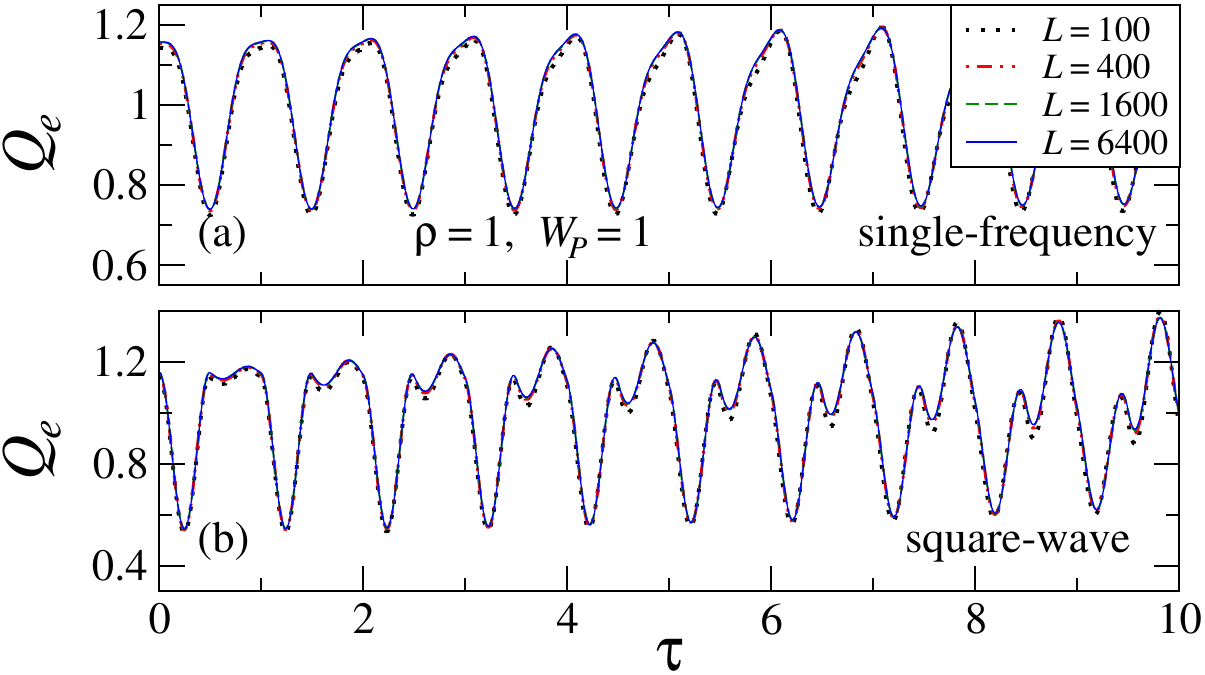}
  \vspace*{2mm} \\
  \includegraphics[width=0.95\columnwidth]{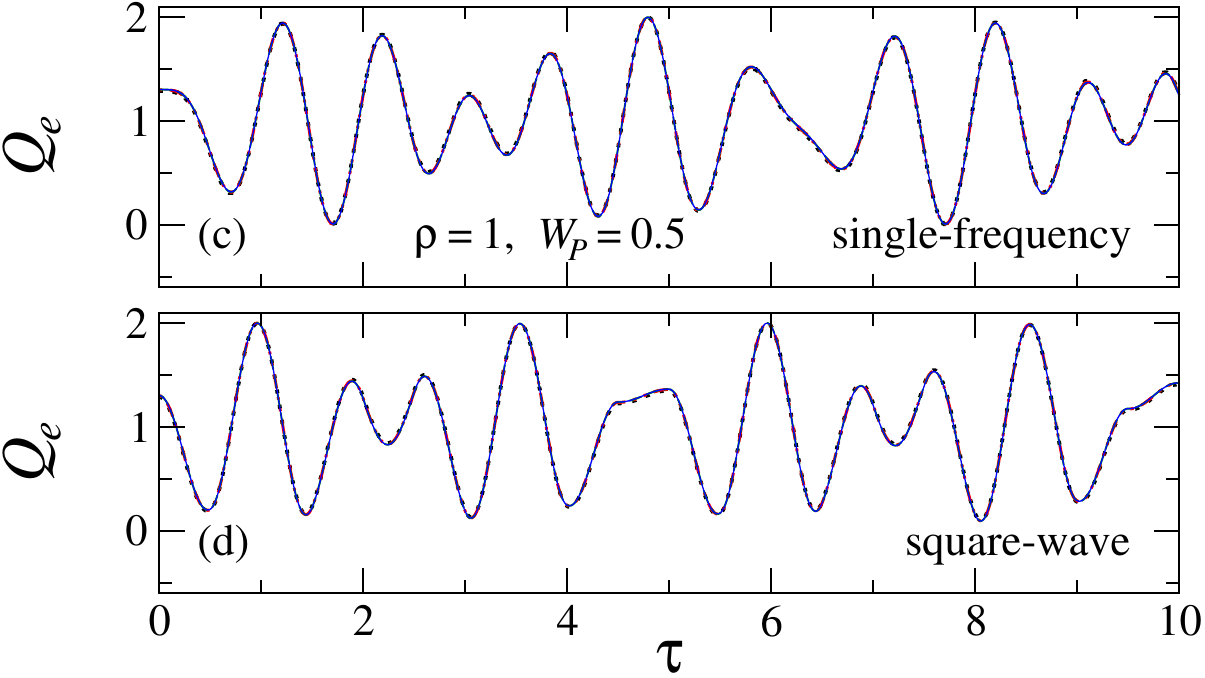}
  \caption{The quantity $Q_e$ defined in Eq.~(\ref{qedef}) vs $\tau=t/P$,
    under single-frequency [(a), (c)] and square-wave [(b), (d)] driving
    of the chemical potential, for $\varrho=1$ and $W_P=1$ [(a), (b)]
    or $0.5$ [(c), (d)]. The various curves, for different system sizes, 
    appear to approach a single scaling curve supporting the dynamic FSS
    in Eq.~(\ref{chifssz2}).}
  \label{Qefss}
\end{figure}

We first consider the quantities $Q_{e/o}$ related to the fermionic
correlators, defined in Eqs.~(\ref{qodef}) and (\ref{qedef}), whose
asymptotic dynamic FSS is expected to be that reported in
Eq.~(\ref{chifssz2}). For this purpose, we have performed computations
at fixed $\varrho = B\,P$ (in particular for $\varrho=1$ and
$1/\sqrt{3}$) and $W_P=P/L$ (in particular, for
$W_P=1,\,0.75,\,0.5,\,0.25$), and various increasing values of $L$, to
observe the convergence of the time dependence of $Q_{e/o}$ in the
dynamic FSS limit as functions of $\tau=t/P$.  A representative sample
of results is shown in Figs.~\ref{Qofss} and \ref{Qefss}.
For both driving protocols and the parameters investigated, the curves
are nearly indistinguishable on the scale of the plots already
for $L \gtrsim 1000$, up to $\tau = 100$. The late-time panels
of Fig.~\ref{Qofss} illustrate this agreement, with only a modest
increase in finite-size effects.
We note a generic oscillatory behavior, which
however does not appear synchronized with the periodic driving.  The
features of the oscillations depend strongly on the scaling
variables, in particular on $W_P$.
The two protocols produce broadly similar oscillatory patterns
for the parameters shown, although their detailed time dependence
and temporal regularity differ quantitatively. In particular,
square-wave driving can produce sharp features at the switching times,
whereas cosine driving yields a smooth evolution.

\begin{figure}[tbp]
  \includegraphics[width=0.95\columnwidth]{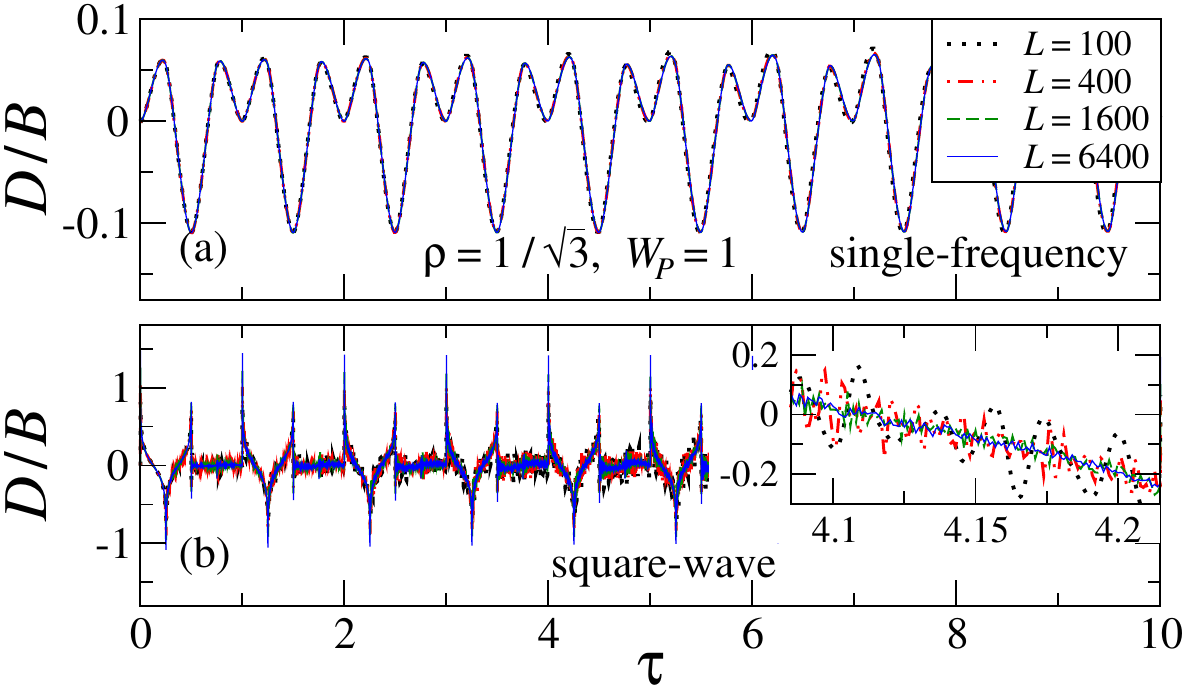}
  \vspace*{2mm} \\
  \includegraphics[width=0.95\columnwidth]{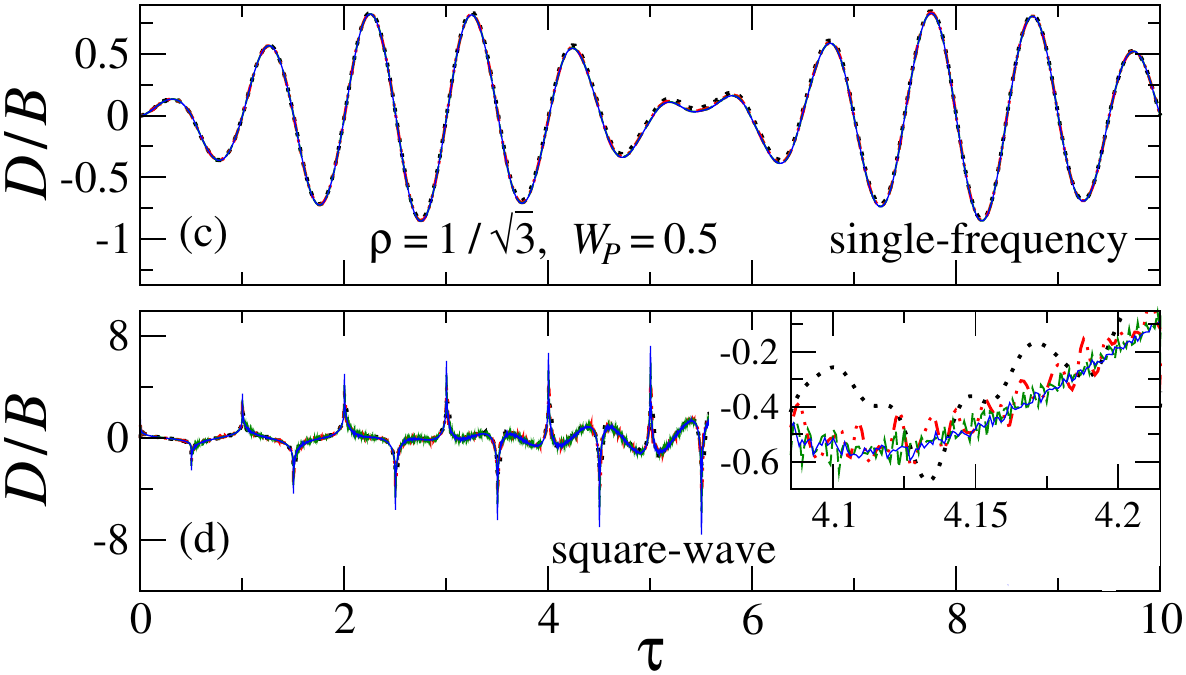} 
  \caption{Dynamic FSS of the subtracted
    particle density $D$ under single-frequency [(a), (c)]
    and square-wave [(b), (d)] driving of the chemical potential.
    We plot $D/B$ versus $\tau=t/P$, for $\varrho = 1/\sqrt{3}$
    and $W_P=1$ [(a), (b)], $0.5$ [(c), (d)]. For square-wave driving,
    convergence toward a common profile, thus supporting
    the scaling description in Eq.~(\ref{pdrfssz2}),
    is observed away from the switching times, where the logarithmic
    jumps of $D/B$ prevent uniform convergence across them.
    The insets in (b) and (d) show the suppression of rapid
    finite-size oscillations with increasing $L$ near $\tau=4.15$.
    Values at switching times refer to the post-switch chemical
    potential.}
  \label{Drhosq3}
\end{figure}

As discussed in Sec.~\ref{Kitdynfsssca}, the dynamic FSS of the
particle density requires the appropriate subtraction defined in
Eq.~(\ref{pdmzz}). The results in Fig.~\ref{Drhosq3} support the dynamic
FSS ansatz~\eqref{pdrfssz2} for smooth single-frequency driving and,
away from the switching times, for square-wave driving.
In the latter case, the discontinuous change of the equilibrium
reference produces jumps in $D$, although the physical particle density
remains continuous, as expressed by Eq.~\eqref{D-jump}. These jumps
should be distinguished from the rapid finite-size oscillations
between switches, whose suppression with increasing $L$ is illustrated
by the insets in panels (b) and (d).

\begin{figure}[tbp]
  \includegraphics[width=0.95\columnwidth]{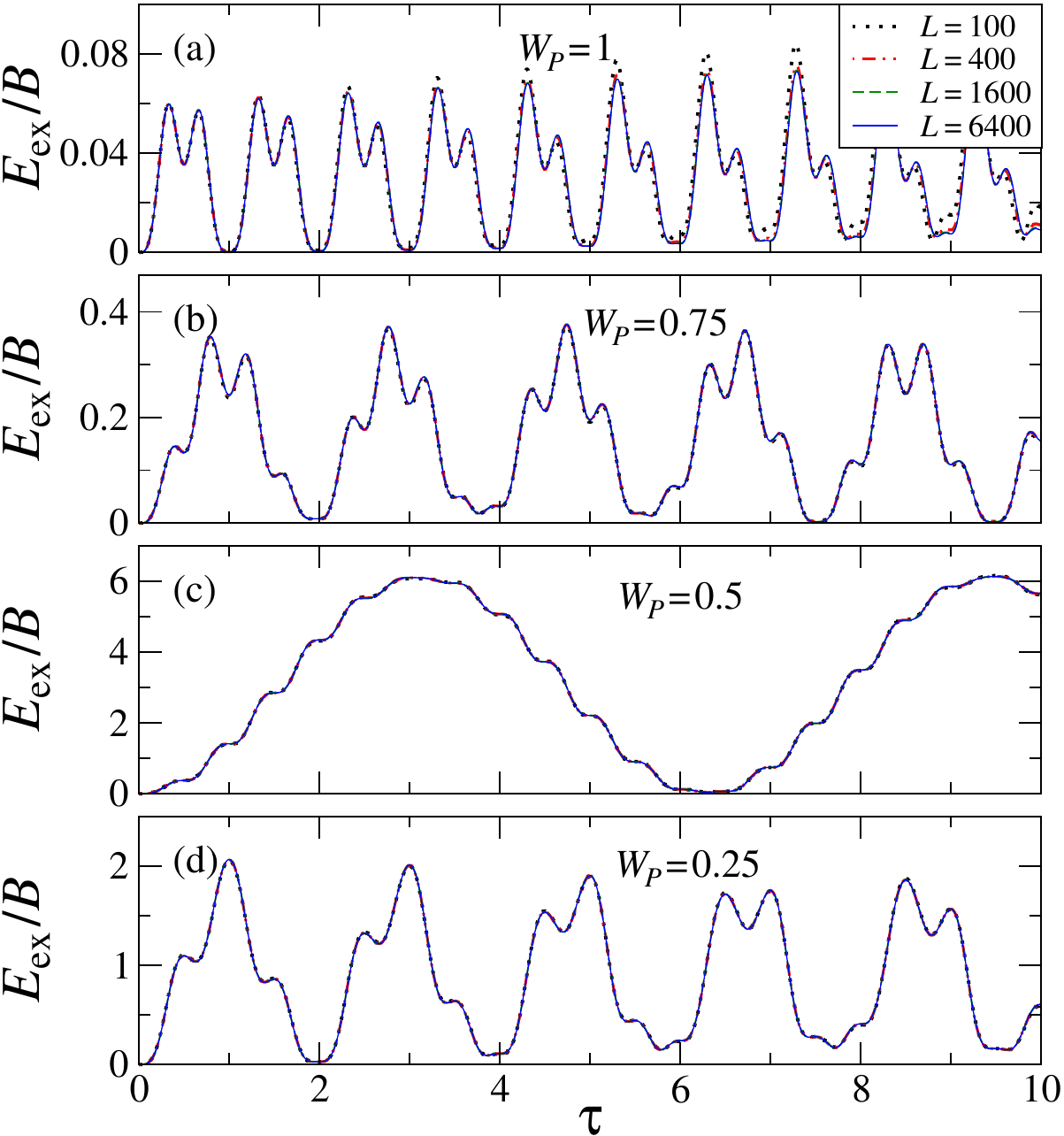}
  \caption{Dynamic FSS of the 
    excess energy $E_{\rm ex}$ under single-frequency periodic driving
    of the chemical potential.  We plot the ratio $E_{\rm ex}/B$ vs
    $\tau=t/P$, for $\varrho = 1$ and various values of $W_P=1$ (a),
    $0.75$ (b), $0.5$ (c), $0.25$ (d).}
  \label{Eexrho1}
\end{figure}

Finally, Fig.~\ref{Eexrho1} shows the excess energy for $\varrho=1$
and several values of $W_P$. In the case of smooth single-frequency
driving, the numerical data nicely support the
asymptotic FSS predicted by Eq.~(\ref{kierfssz}).
For square-wave driving, the corresponding collapse is not observed
(data not shown). This is consistent with the leading logarithmic
contribution generated by the initial sudden quench~\cite{DRV-20,RV-21},
as discussed in App.~\ref{endensque}. Since the Hamiltonian remains
constant during the first half-period, this contribution persists
throughout that interval.

\subsubsection{Dynamic scaling in the thermodynamic limit}
\label{kitthlim}

In quantum Ising chains driven by a periodic transverse field,
certain observables, including the transverse magnetization, tend to
synchronize with the periodic driving in the thermodynamic limit,
developing a large-time stationary dynamics after an initial transient.
Moreover, the system does not generally heat to an
infinite-temperature state, owing to the constraints imposed
by integrability~\cite{RSS-12}. Breaking integrability, e.g.,
making quasiparticles scatter inelastically, can remove these
constraints and, under generic conditions, allow heating toward
an infinite-temperature state.
We have shown that, in the finite-size scaling regime explored above,
the observables display oscillations that are generally
not locked to the driving period over the accessible time windows.
Here, we therefore investigate how the thermodynamic scaling regime
affects their time dependence and approach to long-time behavior.

We first approach the thermodynamic limit by increasing $L$
keeping fixed $B$ and $P$, until the observables become insensitive
to further size increases over the time interval considered.
This is feasible because we can perform computations for very large
chains, with up to $L=O(10^5)$ sites.  The scaling predictions
of Eqs.~(\ref{g12thlim})-(\ref{enethlimkit}),
with the caveats discussed above for square-wave driving,
should then emerge in the limit $B\to 0$ keeping
$\varrho$ and $\tau$ fixed. As already mentioned at the end of
Sec.~\ref{Kitdynsca}, we expect that the approach to the
thermodynamic limit is nonuniform in the rescaled time $\tau$,
requiring $\tau \ll W_P^{-1}$ to suppress finite-size effects.

\begin{figure}[tbp]
  \includegraphics[width=0.95\columnwidth]{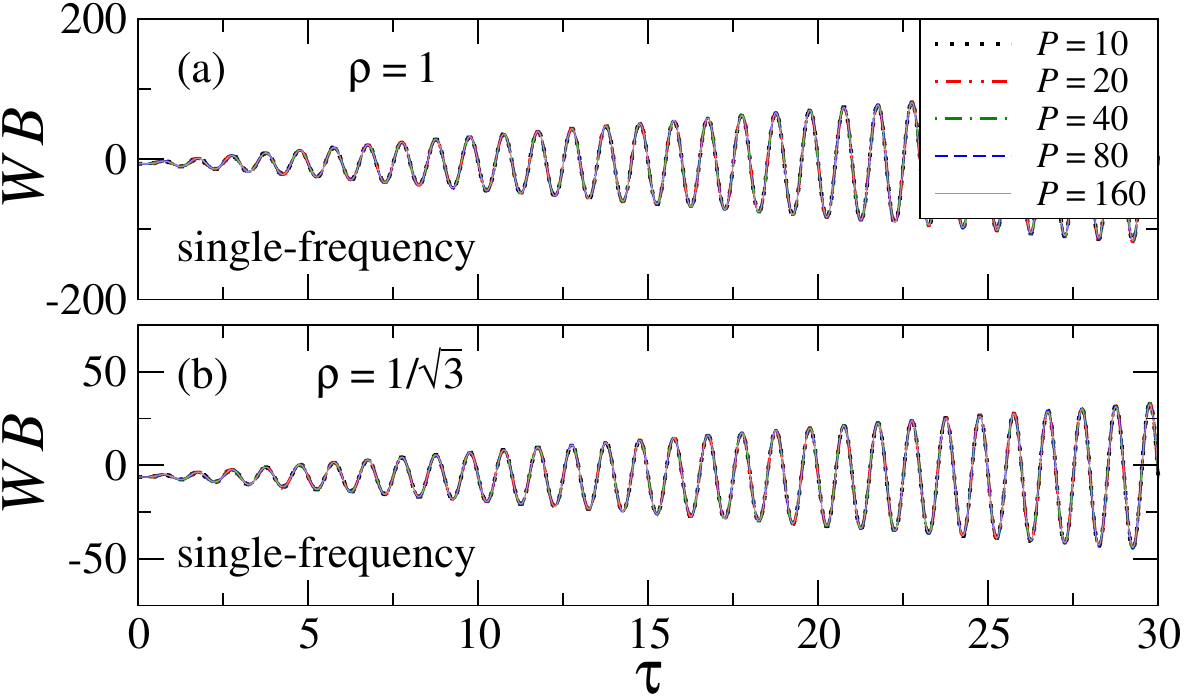}
  \vspace*{2mm} \\
  \includegraphics[width=0.95\columnwidth]{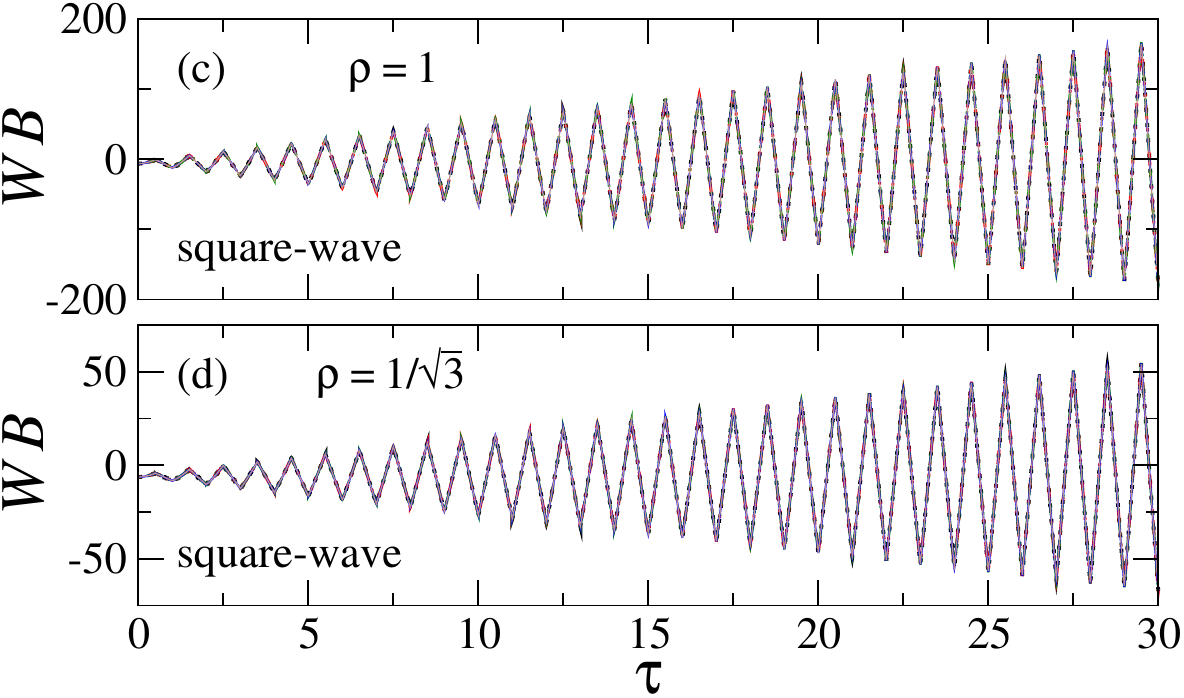}
  \caption{Dynamic scaling of $W$ under
    single-frequency [(a), (b)] and square-wave [(c), (d)]
    driving in the thermodynamic limit. We plot $W\,B$ versus $\tau$ for
    $\varrho=1$ [(a), (c)] and $\varrho=1/\sqrt{3}$ [(b), (d)],
    for various values of the period $P$ (see legend).
    All curves are for $L=204800$ sites, which ensures that
    finite-size effects are negligible on the scale of the figure.}
  \label{Wtl}
\end{figure}  

In Fig.~\ref{Wtl} we show numerical results for the quantity $W=LQ_o$
defined in Eq.~(\ref{wtdef}), whose nontrivial thermodynamic limit
is expected to obey the dynamic scaling law in Eq.~(\ref{wtsca}).
The data support this prediction for both single-frequency
and square-wave drivings.
The scaling curves exhibit oscillations that are locked to the
driving period, corresponding to $\Delta\tau=1$,
while their amplitude appears to grow linearly with $\tau$.
This trend persists at least up to $\tau=100$, as verified
from data extending beyond the range shown in Fig.~\ref{Wtl}.
Thus, the corresponding scaling function
${\cal W}_\infty(\varrho,\tau)$ does not appear to approach a
stationary periodic response within this window. Finally, we note that
the scaling functions obtained for single-frequency and square-wave
drivings are quite similar, although the latter exhibit sharper
oscillations.

\begin{figure}[tbp]
  \includegraphics[width=0.95\columnwidth]{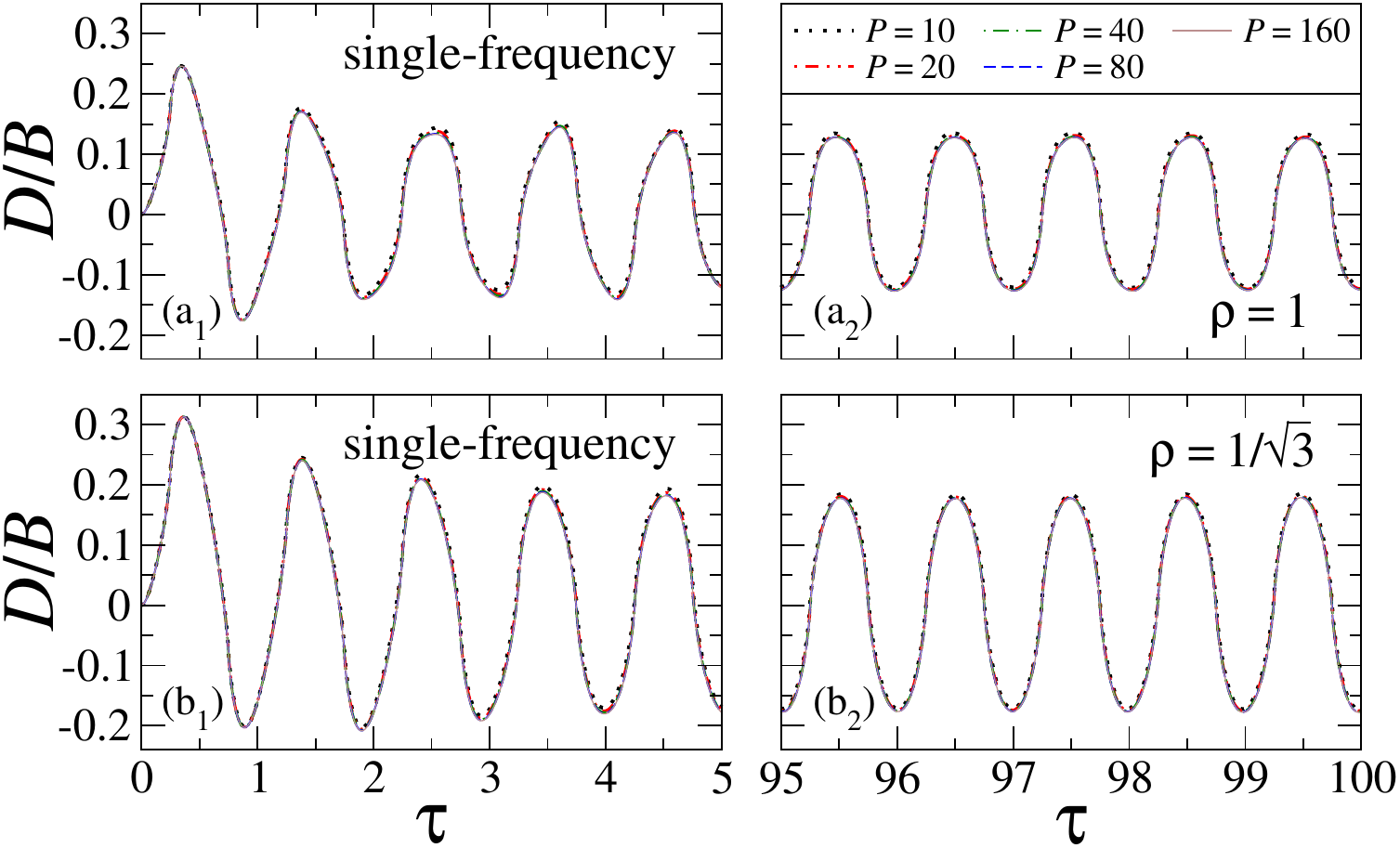}
  \vspace*{3mm} \\
  \includegraphics[width=0.95\columnwidth]{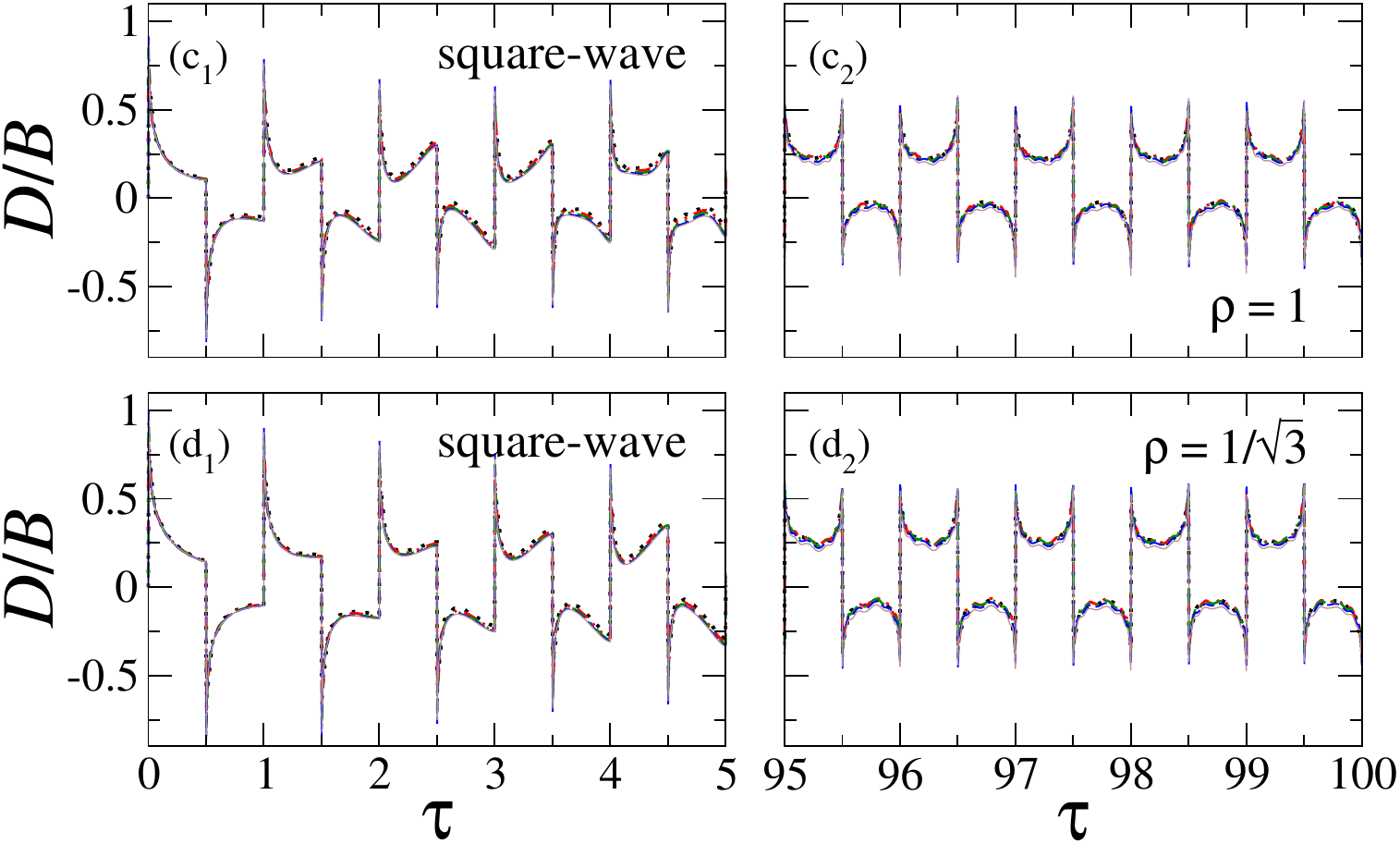}
  \caption{Dynamic scaling of the subtracted particle density
    under single-frequency [rows (a), (b)] and square-wave
    [rows (c), (d)] driving in the thermodynamic limit. We plot $D/B$
    versus $\tau$ for $\varrho=1$ [rows (a), (c)]
    and $\varrho=1/\sqrt{3}$ [rows (b), (d)],
    for several periods $P$. The left and right panels show
    $0 \leq \tau \leq 5$ and $95 \leq \tau \leq 100$, respectively.
    All curves are for $L=204800$ sites, ensuring that finite-size
    effects are negligible on the scale of the figure, even at large times.
    For square-wave driving, values at switching times are evaluated
    immediately after the field change. The scaling comparison
    concerns times away from the switches; the discontinuous equilibrium
    subtraction prevents uniform convergence across them.}
  \label{Dthcos}
\end{figure}  

Figure~\ref{Dthcos} shows the thermodynamic scaling behavior
of the subtracted particle density for both driving protocols.
Square-wave driving produces sharp spikes, whereas cosine
driving yields a smooth response.  At fixed $\varrho$, the curves
approach a common profile as $P$ increases, supporting the predicted
thermodynamic scaling; for square-wave driving, this comparison
applies away from the switching times.
After an initial transient, the curves display an approximately
periodic pattern on the scale of the figure. Small cycle-to-cycle
variations nevertheless persist, as quantified below for cosine driving.

\begin{figure}[tbp]
  \includegraphics[width=0.95\columnwidth]{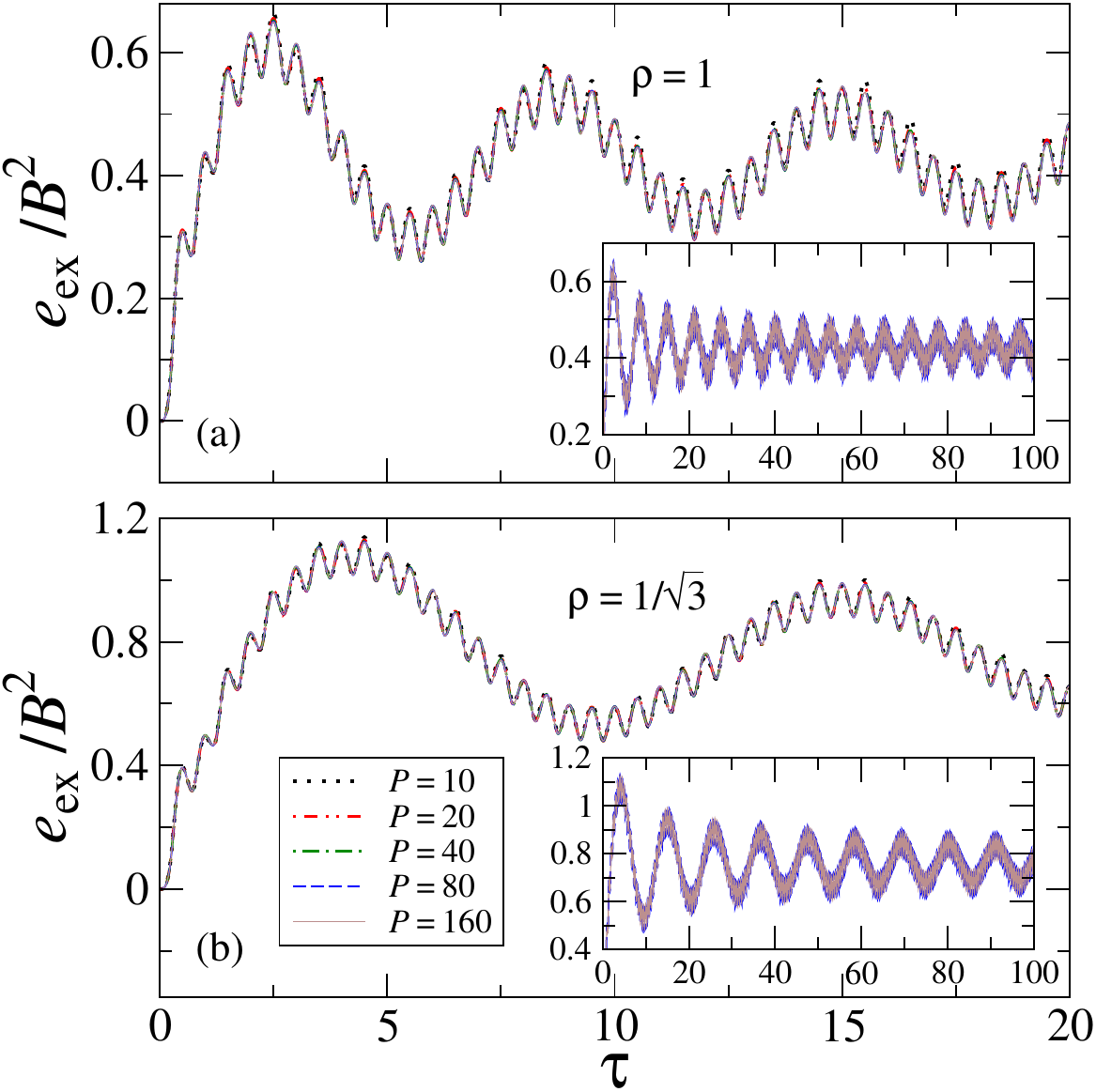}
  \caption{Dynamic scaling of the excess-energy density under
    single-frequency driving in the thermodynamic limit.
    We plot $e_{\rm ex}/B^2$ versus $\tau$ for $\varrho=1$ (a) and
    $1/\sqrt{3}$ (b), for several periods $P$.
    The main frames are for $\tau \in [0,20]$, while the two
    insets report the same curves at the two largest $P=80, \, 160$
    in the interval $\tau \in [0,100]$.  All curves are for
    $L=204800$ sites, with finite-size effects negligible
    on the scale of the figure.}
  \label{Eextlcos}
\end{figure}  

Figure~\ref{Eextlcos} shows the thermodynamic scaling of the
excess-energy density under cosine driving, as predicted
by Eq.~\eqref{enethlimkit}, for times up to $\tau = 100$.
As a consistency check, we numerically integrated the right-hand
side of Eq.~\eqref{eex-D}, using the density data shown
in the single-frequency-driving panels of Fig.~\ref{Dthcos},
and compared the result with the independently
computed excess-energy density shown in Fig.~\ref{Eextlcos}.
For all cosine-driving parameter sets considered, the maximum
absolute discrepancy in the rescaled excess-energy density
remains below $2 \times 10^{-6}$ over $0 \leq \tau \leq 100$.
The excess-energy density exhibits slow modulations together
with faster oscillations. Although these slow modulations are more conspicuous
than the cycle-to-cycle variations of the density in Fig.~\ref{Dthcos},
the two behaviors are connected by Eq.~\eqref{eex-D}.
Multiplication by the driving sine mixes frequencies, while time
integration gives greater weight to low-frequency contributions.
Small deviations of the density from a periodic response
can therefore produce visible slow modulations of the excess energy.

\begin{figure}[tbp]
  \includegraphics[width=0.95\columnwidth]{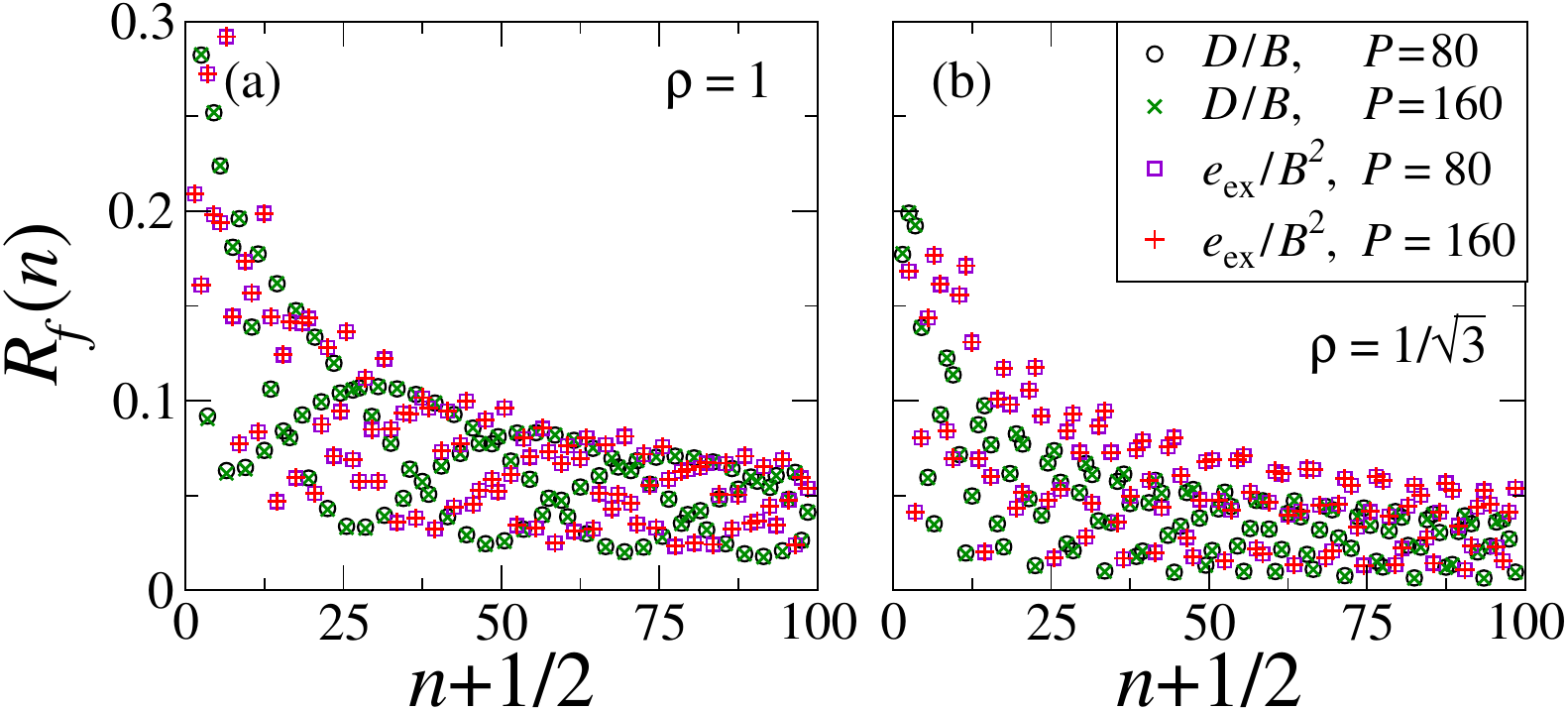}
  \caption{Relative cycle-to-cycle deviations $R_f(n)$
    for the subtracted density $f = D/B$ and excess-energy density
    $f = e_{\rm ex}/B^2$  under cosine driving, plotted versus
    the midpoint $n+1/2$ of the first cycle in each pair.
    Panels (a) and (b) correspond to $\varrho=1$
    and $\varrho=1/\sqrt{3}$, respectively.
    Black and red symbols denote $P=80$ and $160$, respectively;
    symbol shapes distinguish the observables (see legend).
    For each observable, the results for the two periods are nearly
    indistinguishable on the scale of the figure.
    All data are for $L=204800$.  Each point compares two consecutive
    cycles, normalized by the root-mean-square signal in the first
    cycle without subtracting its mean.}
  \label{fig:cycle_rms}
\end{figure}  

To compare the two observables quantitatively, we introduce
the relative cycle-to-cycle deviation
\begin{equation}
  R_f(n)= \left[
    \frac{\int_n^{n+1} [f(\tau+1)-f(\tau)]^2 \, d\tau}
         {\int_n^{n+1} f(\tau)^2 \, d\tau} \right]^{1/2},
\end{equation}
where $f=D/B$ or $f=e_{\rm ex}/B^2$. Figure~\ref{fig:cycle_rms} reports
$R_f(n)$ versus $n+1/2$ for cosine driving.
Both observables retain measurable deviations from a period-one
response, with an overall decreasing envelope. The results for $P=80$
and $160$, at fixed $\varrho$, are nearly indistinguishable.
Thus, the different visual appearance of the cosine-driving
panels of Figs.~\ref{Dthcos} and~\ref{Eextlcos} does not imply
a qualitative distinction
between a synchronized density and an unsynchronized excess energy.
Rather, both exhibit residual cycle-to-cycle variations over
the investigated time window. Their decrease suggests an approach
toward a periodic response, without establishing
the ultimate long-time behavior.

\begin{figure}[tbp]
  \includegraphics[width=0.95\columnwidth]{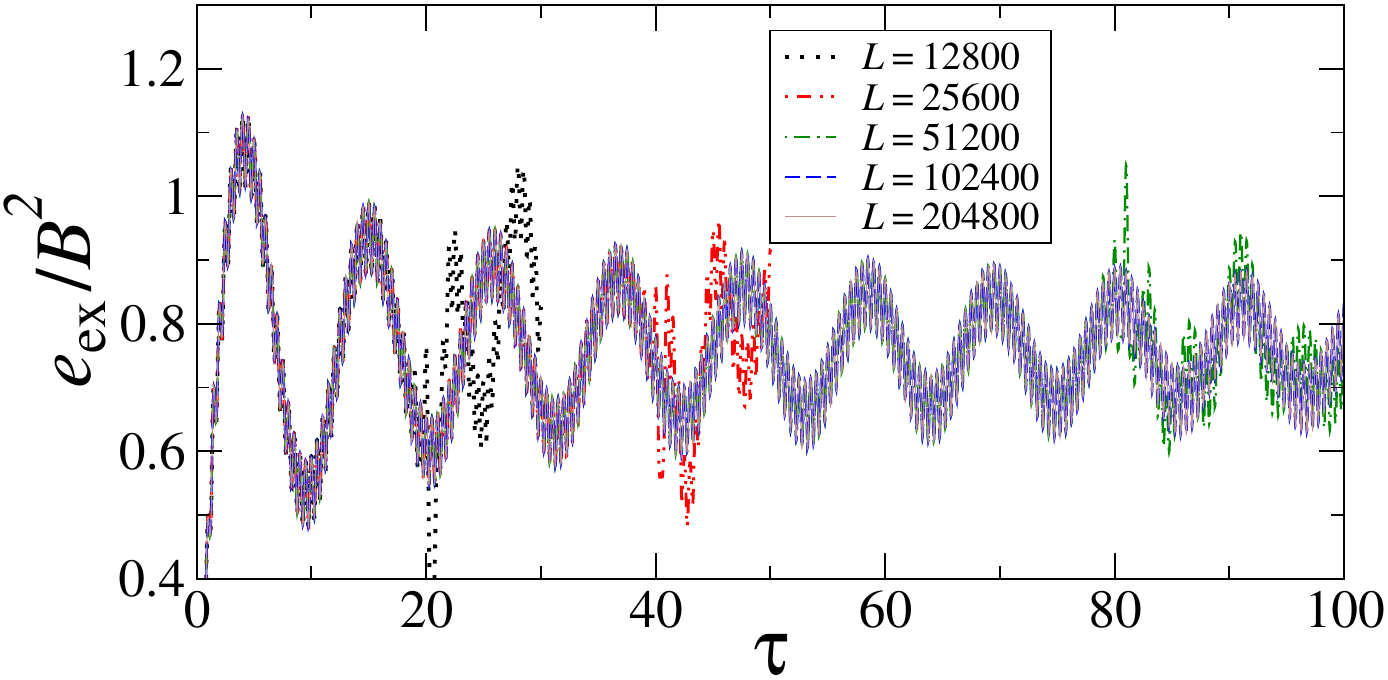}
  \caption{Approach to the thermodynamic limit of the excess-energy
    density under cosine driving. We plot $e_{\rm ex}/B^2$
    versus $\tau$ for $\varrho=1/\sqrt{3}$, $P=160$, and several
    system sizes $L$.  Note that the onset of finite-size
    effects shifts to later times, approximately proportional to $L$.}
  \label{Eextlcosl}
\end{figure}  

Finally, we come back to the thermodynamic-limit issue, illustrating
how it is approached. For this purpose, in Fig.~\ref{Eextlcosl} we
show some results for the excess-energy density at fixed $B$ and $P$,
for increasing system sizes $L$. The curves agree within the graphical
resolution at early times, while finite-size deviations appear
progressively later as $L$ increases, with an onset time approximately
proportional to $L$. This behavior clearly demonstrates the
nonuniformity of convergence in time discussed in
Sec.~\ref{Kitdynsca}, consistent with the condition
$\tau W_P=t/L\ll1$ for suppressing finite-size effects.

\section{Conclusions}
\label{conclu}

We investigate the out-of-equilibrium scaling behaviors induced by
periodic driving of Hamiltonian parameters in quantum many-body
systems near continuous quantum phase transitions. Specifically, we
consider two paradigmatic one-dimensional quantum many-body systems,
the quantum Ising chain and the Kitaev fermionic wire, which are
related by the standard nonlocal Jordan–Wigner transformation. We
focus on two classes of driving protocols, in which either the
longitudinal or the transverse field drive the quantum Ising system
across the critical point during each cycle.

We analyze the time-dependence of several different observables using
an RG framework, determining their dynamic scaling behavior both for
finite-size systems and in the infinite-volume limit.  While the
underlying scaling arguments are general and not specific to the
models considered, the quantitative results depend on the particular
system and, more specifically, the resulting scaling functions also
depend on the driving waveform and on the initial state.
We distinguish between protocols that break the global symmetry, such
as periodic longitudinal-field driving, and those that preserve it,
such as periodic transverse-field driving--- equivalently, periodic
chemical-potential driving in the Kitaev wire. These two classes of
driving protocols exhibit distinct dynamic scaling behaviors,
characterized by different power-law exponents determined by the RG
dimensions of the corresponding perturbations.

We test these predictions numerically in Ising chains driven by a
periodic square-wave longitudinal field and in Kitaev wires with both
cosine and square-wave modulations of the chemical potential.  Our
results confirm dynamic FSS for longitudinally driven Ising chains and
both finite-size and thermodynamic-limit scaling for Kitaev wires. The
approach to the scaling regime depends on the driving parameters and the time
interval considered, requiring progressively larger systems to
resolve the dynamics at longer times.  The comparison between smooth
and discontinuous driving also reveals important qualifications to the
scaling predictions.  For square-wave driving of the Kitaev wire, the
initial quench generates a logarithmic contribution that prevents the
excess energy from obeying the proposed pure power-law scaling.  The
subtracted density exhibits jumps due to the sudden change of the
instantaneous equilibrium reference value, so that scaling away from the
switching times must be distinguished from nonuniform convergence in their
vicinity.

A notable difference between the FSS and thermodynamic-limit scaling
regimes concerns the relation between the observable dynamics and the
driving period.  In the FSS regime, the observables generally display
oscillations that are not locked to the driving period over the
accessible time windows. Instead, in the thermodynamic scaling regime,
the density exhibits an approximately periodic pattern after an
initial transient.  Under cosine driving, however, a quantitative
comparison of successive cycles reveals residual variations in both
the density and the excess energy, with an overall decreasing
envelope.  The exact density–energy relation explains how small
cycle-to-cycle variations of the particle density can produce more
conspicuous slow modulations of the excess energy.  These results
suggest an approach toward a periodic response, but do not establish
its asymptotic form.  Meanwhile, $W$ exhibits period-locked
oscillations with a growing envelope, illustrating that oscillations
at the driving period do not by themselves imply convergence to a
stationary periodic response.  The possible emergence of such a
response in the thermodynamic limit connects with earlier studies of
integrable driven chains~\cite{RSS-12}, extending this discussion to
the critical scaling regime.

An interesting perspective would be to investigate spatially localized
periodic driving associated with single-site defects.  A corresponding
dynamic scaling framework could build on previous studies of localized
perturbations near quantum transitions (see, e.g.,
Refs.~\cite{PAW-17,PRV-18d,FRV-22eq,FRV-22}).  Such protocols would
allow one to explore how the driving amplitude and period control the
spatially inhomogeneous response and the propagation of correlations
and energy away from the driven region.

Another natural extension concerns periodic driving across first-order
quantum transitions~\cite{Pfleiderer-05,RV-21,PV-24}.  Existing
studies have established dynamic scaling behaviors after sudden
quenches and during slow passages across these transitions (see, e.g.,
Refs.~\cite{PRV-18,PRV-20,TS-23,PV-24,PRV-25}).  Extending these
scaling frameworks to periodic protocols would allow one to
investigate the effects of repeated crossings, both in finite-size
systems and in the thermodynamic limit.  Some evidence of scaling
under periodic driving has been already reported in Ref.~\cite{TS-23}.

It would also be useful to assess the robustness of the observed
scaling behaviors to weak integrability-breaking interactions
and to coupling with an environment. The former would clarify
the time scales over which the dynamics of the quadratic Kitaev
model remains representative of interacting systems. Dissipation,
on the other hand, introduces a competition between coherent
evolution and relaxation that may qualitatively change the
long-time response, as illustrated by periodically driven two-level
quantum systems, where it may induce synchronization at long times
(see, e.g., Refs.~\cite{HG-10,RPBF-11}).
Thus, dissipative periodically driven
quantum many-body systems may exhibit behaviors similar to those
observed in periodically driven classical spin systems with
dissipative relaxational dynamics~\cite{PV-26-per}.
A Lindblad master-equation description~\cite{Lindblad-76,GKS-76}
(see also Refs.~\cite{BP-book,RH-book,SBD-16})
would provide a natural framework for investigating whether
this competition leads to asymptotically periodic responses and how
it modifies dynamic scaling near the quantum transition.

\appendix

\section{Energy density after a quench}
\label{endensque}

We consider a sudden quench of the transverse field $g$, in the
absence of longitudinal field ($h=0$). The system is initially prepared
in the ground state at $g=g_i$. Then, at $t=0$, the transverse field
is suddenly changed to $g$.
The energy evaluated with the post-quench Hamiltonian is conserved
for $t > 0$, i.e.,
\begin{eqnarray}
  E(t) &=& \langle \Psi(t) | \hat H(g) | \Psi(t)\rangle \nonumber\\
  &=& \langle \Phi_0(g_i) | \hat H(g) | \Phi_0(g_i)\rangle \equiv  L e_q.
  \label{et0+}
\end{eqnarray}
In the following, all quantities are evaluated in the thermodynamic limit.
We retain the explicit size argument in the ground-state energy density
$e_0(g,\infty)$ and the transverse magnetization $U_0(g,\infty)$,
to distinguish these equilibrium quantities from their
finite-size counterparts.

To evaluate the post-quench energy density in the thermodynamic limit,
we use the identity
\begin{equation}
  \hat H(g) = \hat H(g_i) - (g-g_i) L \hat U,
\end{equation}
where $\hat U = L^{-1} \sum_x \hat\sigma_x^{(3)}$ is the transverse
magnetization operator,
and the exact expressions for the ground-state energy density
and transverse magnetization~\cite{LSM-61,Katsura-62,Pfeuty-70,BMD-70}:
\begin{eqnarray}
  e_0(g,\infty) &=& \lim_{L\to \infty} L^{-1}
  \langle \Phi_0(g) | \hat H(g) | \Phi_0(g)\rangle \nonumber\\
  &=&-\frac{1}{\pi} \int_0^\pi dk\,\sqrt{1 + g^2 - 2g \cos k}  , \\
  U_0(g,\infty) &=& \lim_{L\to \infty}
  \langle \Phi_0(g) | \hat U | \Phi_0(g)\rangle \nonumber\\
  &=& \frac{1}{\pi} \int_0^\pi dk\,
{ g - \cos k\over  \sqrt{1 + g^2 - 2g \cos k}} . \label{Ugval}
\end{eqnarray}
The post-quench energy density is then given by
\begin{equation}
  e_q(g_i,g) = -\frac{1}{\pi} \int_0^\pi dk\,
  {1 + g_i g - (g_i+g)\cos k\over
    \sqrt{1 + g_i^2 - 2g_i \cos k}}.
  \label{eq-ex}
\end{equation}
The work per site performed during the quench and the excess
energy density are, respectively:
\begin{eqnarray}
  w(g_i,g) &=& e_q(g_i,g) - e_0(g_i,\infty), \label{work}\\ 
  e_{\rm ex}(g_i,g) &=& e_q(g_i,g) - e_0(g,\infty) \ge 0. \label{exene} 
\end{eqnarray} 

We now consider a symmetric quench across the critical point,
from $g_i = g_c + \delta$ to $g = g_c - \delta$, with $\delta > 0$.
Expanding Eq.~\eqref{eq-ex} and subtracting the final ground-state
energy density, we obtain, as $\delta \to 0^+$,
\begin{eqnarray}
  \tilde e_{\rm ex}(\delta) &\equiv& e_{\rm ex}(g_c+\delta,g_c-\delta) \nonumber\\
  &=& - \frac{2}{\pi} \delta^2 \ln \delta + \frac{6\ln 2 - 2}{\pi}
\delta^2 + o(\delta^2).  \label{eexasy}
\end{eqnarray}
The leading term contains a multiplicative logarithm and therefore
does not obey the pure power-law scaling prediction
\begin{equation}
\tilde e_{\rm ex}(\delta) \sim \delta^{\nu(1+z)} = \delta^2.
\end{equation}
In particular, the ratio of the excess energy density to $\delta^2$
diverges logarithmically as $\delta \to 0^+$.

The quench considered above coincides with the initial step of the
square-wave transverse-field protocol introduced in Sec.~\ref{traper},
with amplitude $B_g = \delta$. Since the Hamiltonian remains constant
during the first half-period, the excess energy retains its post-quench
value for $0 < t < P/2$. The logarithmic contribution therefore prevents
the proposed pure power-law scaling from holding throughout this interval.

The same conclusion applies to the square-wave chemical-potential protocol
of the Kitaev chain. Under the mapping $g = -\mu/2$, its initial quench
from $\mu_c + B$ to $\mu_c - B$ corresponds to a transverse-field quench
from $g_c - B/2$ to $g_c + B/2$. This reverses the direction of the quench
considered above. Interchanging the initial and final fields
in Eq.~\eqref{eq-ex} yields the same leading logarithmic term,
with $\delta = B/2$. Consequently, the Kitaev excess-energy density
contains a leading contribution $-(B^2 \ln B)/(2\pi)$, which is not
removed by subtracting the final ground-state energy.
This argument concerns the initial discontinuous quench of the
square-wave protocol and does not imply a breakdown of the
power-law scaling observed for smooth cosine driving.

\end{document}